\documentclass[aip,
               preprint,
               amsmath,
               superscriptaddress,
               floatfix,
               nobibnotes,
              ]{revtex4-2}
              
\usepackage{amsfonts}
\usepackage{bm}
\usepackage[dvips]{graphicx}
\usepackage{color}
\usepackage[utf8]{inputenc}
\usepackage[flushleft]{caption}
\usepackage{subfigure}
\usepackage{rotating}
\usepackage{txfonts}
\usepackage{booktabs}
\usepackage[unicode]{hyperref}
\usepackage{xcolor}
\usepackage{indentfirst}
\usepackage{mathrsfs}  

\DeclareMathAlphabet{\mathpzc}{OT1}{pzc}{m}{it}

\def\figfoot{CPD-SVR, Meng}

\newcommand{\figcaption}[2]{
    \noindent {\bf Figure \ref{#1}:} #2
    \vspace{1cm}
}

\begin{document}


\title{Canonical-Polyadic-Decomposition of the Potential Energy
Surface Fitted by Warm-Started Support Vector Regression}

\author{Zekai Miao}
 \affiliation{Department of Chemistry,
              Northwestern Polytechnical University,
              West Youyi Road 127, 710072 Xi'an,
              China}
\affiliation{Department of Computer Science and Technology,
             University of Cambridge,
             15 JJ Thomson Avenue,
             CB3 0FD Cambridge,
             United Kingdom}              
              
\author{Xingyu Zhang}
 \affiliation{Department of Chemistry,
              Northwestern Polytechnical University,
              West Youyi Road 127, 710072 Xi'an,
              China}

\author{Qingfei Song}
 \affiliation{Department of Chemistry,
              Northwestern Polytechnical University,
              West Youyi Road 127, 710072 Xi'an,
              China}

\author{Qingyong Meng}
 \email{qingyong.meng@nwpu.edu.cn}
  \affiliation{Department of Chemistry,
               Northwestern Polytechnical University,
               West Youyi Road 127, 710072 Xi'an,
               China}

\date{\today}


\begin{abstract}

~~~\\
{\bf Abstract}:
In this work, we propose a decoupled support vector regression (SVR)
approach for direct canonical polyadic decomposition (CPD) of a potential
energy surface (PES) through a set of discrete training energy data.
This approach, denoted by CPD-SVR, is able to directly construct the
PES in CPD with a more compressed form than previously developed
Gaussian process regression (GPR) for CPD, denoted by CPD-GRP ({\it J.
Phys. Chem. Lett.} {\bf 13} (2022), 11128). Similar to CPD-GPR, the
present CPD-SVR method requires the multi-dimension kernel function
in a product of a series of one-dimensional functions. We shall show
that, only a small set of support vectors play a role in SVR prediction
making CPD-SVR predict lower-rank CPD than CPD-GPR. To save computational
cost in determining support vectors, we propose a warm-started (ws)
algorithm where a pre-existed crude PES is employed to classify the
training data. With the warm-started algorithm, the present CPD-SVR
approach is extended to the CPD-ws-SVR approach. Then, we test CPD-ws-SVR
and compare it with CPD-GPR through constructions and applications of
the PESs of H + H$_2$, H$_2$ + H$_2$, and H$_2$/Cu(111). To this end,
the training data are computed by existed PESs. Calculations on H + H$_2$
predict a good agreement of dynamics results among various CPD forms, which
are constructed through different approaches.
{\it Support Vector Regression}; {\it Quantum Dynamics}; {\it Low Rank}

\end{abstract}

\maketitle

\section{Introduction\label{sec:intro}}

In propagating nuclear wave function with the grid-based representation
\cite{mil90:245,mey93:141}, the whole configurational space is represented
by a given set of grids. In this context, an $f$-dimension potential
energy surface (PES) as a continuous function of coordinates,
$V(q^{(1)},\cdots,q^{(f)})$, has to be (approximately) represented in
a tensor form $V(q^{(1)}_{i_1},\cdots,q^{(f)}_{i_f})=V_{i_1,\cdots,i_f}=V_I$,
where $\{q^{(\kappa)}\}_{\kappa=1}^f$ is the set of $f$ coordinates
and $I=\{i_{\kappa}\}_{\kappa=1}^f$ denotes a set of tensor indices.
Moreover, in modern approaches to propagate nuclear wave function,
such as multi-configurational time-dependent Hartree (MCTDH) method
\cite{mey90:73,man92:3199,bec00:1} and its multi-layer version (ML-MCTDH)
\cite{man08:164116,dor08:224109,ven11:044135}, the nuclear wave function
must be expanded as a series of products of few-particle functions. In
Heidelberg version of MCTDH and ML-MCTDH \cite{mey90:73,man92:3199,bec00:1,ven11:044135},
the potential tensor $V_I$ has to be accordingly transferred to a
sum-of-products (SOP) form
\cite{jae95:5605,jae96:7974,jae98:3772,pel13:014108,pel14:42,ott14:014106,ott17:116,sch17:064105}
or canonical polyadic decomposition (CPD) form \cite{sch20:024108,men21:2702,shi23:194102}.
The latter is a special class of SOP. In this work, since the SOP and
CPD forms play the above important role in wave function propagations,
we propose a new approach to build the PES in the CPD form through
support vector regression (SVR). It was developed as one of data-driven
techniques with sparse kernel function space and was recently suggested
\cite{son22:1983} as a possible approach in building the PES in general
form.

Usually, the potential function in the SOP/CPD form is constructed using
an existing analytical PES through an appropriate algorithm leading to
a two-step scheme (see counterclockwise steps in Figure \ref{fig:methods}).
First, the analytical PES is constructed through a
database. Second, the PES in the SOP/CPD form is re-fitted.
In 1990s, Meyer and co-workers \cite{jae95:5605,jae96:7974,jae98:3772}
proposed the POTFIT (means potential re-fitting) algorithm to construct
the SOP form on grids whose total amount must be less than $2\times10^9$.
For high-dimension models, the multi-grid POTFIT (MGPF) \cite{pel13:014108,pel14:42},
multi-layer POTFIT (MLPF) \cite{ott14:014106,ott17:116}, Monte-Carlo
POTFIT (MCPF) \cite{sch17:064105} algorithms have been designed and
implemented. Schr{\"o}der \cite{sch20:024108} found that the PES in
the CPD form is effective for high-dimension propagation. Employing
the Monte Carlo approach, the MCCPD algorithm \cite{sch20:024108} was
proposed to construct the CPD form. Later, the MCCPD method in conjugation
with $21$-dimension (21D) ML-MCTDH calculations were launched \cite{men21:2702}
proving the power of MCCPD as well as the CPD form.
Recently, a 75D PES of the hydrogen atom scattering off non-rigid
graphene surface \cite{shi23:194102} was re-fitted in the CPD form
by MCCPD further showing its power.
Although the above algorithms for the two-step scheme were proved to
be powerful
\cite{jae95:5605,jae96:7974,jae98:3772,pel13:014108,pel14:42,ott14:014106,ott17:116,sch17:064105,sch20:024108,men21:2702}
in re-fitting the PES in either the SOP/CPD form,
direct construction of SOP or CPD potential function through discrete
energy data must be developed leading to a one-step scheme. As shown
by clockwise step of Figure \ref{fig:methods},
the one-step scheme avoids constructing the PES in general form and
thus saves computational cost because constructions of both general
form and SOP/CPD form are very expensive.
In 2006, Manzhos and Carrington Jr. \cite{man06:194105} used exponential
activation function to construct a neural-network (NN) function to
represent potential functions in the CPD form. In 2014, Koch and Zhang
\cite{koc14:021101} found another kind of single-layer NN function for
constructing the CPD potential and proposed the SOP-NN method.
In addition, Pel{\'a}ez and co-workers \cite{pan20:234110,nad23:114109}
proposed a SOP or CPD form of finite basis representation (FBR), denoted
by SOP-FBR or CP-FBR. By both of them, a guess of initial SOP/CPD functional
form is optimized to yield either SOP or CPD form through existing potential
function.

In 2022, a direct construction of the CPD form, called CPD-GPR
\cite{son22:11128}, was proposed on the basis of Gaussian process
regression (GPR) \cite{son20:134309,son22:1983}. Recently, combining
several coordinates with strong couplings into one logical coordinate,
the mode-combination (mc) version of CPD-GPR was proposed \cite{son24:597},
denoted by CPD-mc-GPR. Employing CPD-mc-GPR, a 9D PES of OH + HO$_2$
was directly constructed in mode-combined CPD form \cite{son24:597}.
This construction under the one-step scheme opening a new door to
build the PES of realistic reaction in CPD tensor form. However, both
CPD-GPR and CPD-mc-GPR produce number of decomposition terms, called
rank of CPD, is equal to number of the training energy data
\cite{son22:11128,son24:597}. Comparing with MCCPD \cite{sch20:024108}
that gives the CPD rank of $10^2\sim10^3$, a typical rank of
$10^3\sim10^4$ provided by CPD-mc-GPR largely reduces the efficiency
of subsequent quantum dynamics calculation \cite{son24:597}.
To overcome this problem, in this work noting its sparse kernel
feature, the SVR method is further developed leading to a new
implementation of the one-step scheme, called CPD-SVR. Because
prediction of SVR depends on the number of support vectors which
is usually smaller than that of training data \cite{son22:1983},
the present CPD-SVR method is able to produce the CPD form with
smaller rank. Furthermore, SVR consumes lots of computational cost in
determining support vectors, a warm-started (ws) algorithm is proposed
in this work to improve efficiency leading to CPD-ws-SVR. Essentially,
the present CPD-ws-SVR is a kernel regression with a kernel function
in product form, while a selection algorithm for choosing support vectors
is employed to accelerate convergence. In this work, we test the present
CPD-ws-SVR method and compare it with CPD-GPR by building the PES of H +
H$_2$. The CPD form is also constructed by the POTFIT and MCCPD methods
to show the power of CPD-ws-SVR by performing MCTDH calculations.

The rest of this paper is organized as follows. In Section \ref{sec:theo},
we shall describe the theoretical framework of CPD-ws-SVR. Section
\ref{sec:results} presents details of test calculations. Finally,
Section \ref{sec:con} concludes with a summary.

\section{Theoretical Framework\label{sec:theo}}
\subsection{The CPD-SVR Method\label{sec:sop-gpr}}

First of all, we have to briefly describe concept of the CPD form. For
an $f$-dimensional potential function $V(q^{(1)},\cdots,q^{(f)})$, its
CPD form on a set of grids $I=\{i_{\kappa}\}_{\kappa=1}^f$ is given by
\cite{sch20:024108,men21:2702,son22:11128}
\begin{equation}
V\big(q^{(1)}_{i_1},\cdots,q^{(f)}_{i_f}\big)\simeq
V^{(\mathrm{CPD})}\big(q^{(1)}_{i_1},\cdots,q^{(f)}_{i_f}\big)=
V_I^{\mathrm{(CPD)}}=
\sum_r^Rc_r\prod_{\kappa}^fv_{r,i_{\kappa}}^{(\kappa)}
=\sum_r^Rc_r\Omega_{r,I},
\label{eq:cpd-potential-form-02}
\end{equation}
where $R$ is the expansion order, that is the CPD rank, while the
product $\Omega_{r,I}=\prod_{\kappa}^fv_{r,i_{\kappa}}^{(\kappa)}$
is introduced for clarity. The normalized functions
$v_r^{(\kappa)}(q^{(\kappa)})$ are called the single-particle potentials
(SPP), exclusively depend on only one variable. Here, we denote that
$v_{r,i_{\kappa}}^{(\kappa)}=v_{r}^{(\kappa)}(q^{(\kappa)}_{i_\kappa})$
for the $i_{\kappa}$-th grid of the $\kappa$-th degree of freedom (DOF).
Having $V_I^{(\mathrm{CPD})}$ in Equation \eqref{eq:cpd-potential-form-02},
the remaining task is to find both the expansion functions $\Omega_{r,I}$
and the coefficients $c_r$. This motivates the present developments of
CPD-SVR and its warm-started version (CPD-ws-SVR).

Before giving CPD-SVR and CPD-ws-SVR, a brief description of SVR is
necessary. We shall return to the warm-started SVR (ws-SVR) later. To
construct the $f$-dimensional PES, a total of $n$ energy points
$\{\mathbf{X},\mathbf{E}\}=\{\mathbf{X}_j,E_j\}_{j=1}^n$, must be
computed in advance by {\it ab initio}. Matrices $\mathbf{X}$ and
$\mathbf{E}$ are $f\times n$ coordinate matrix and $n\times1$ energy
matrix, respectively. Then, a regression model that reproduces the
energy at $f\times1$ coordinate vector $\mathbf{x}$, denoted by
$y(\mathbf{x})$, is optimized through $\{\mathbf{X},\mathbf{E}\}$. In
general, the SVR prediction function \cite{son22:1983},
\begin{equation}
y(\mathbf{x})=\boldsymbol{\phi}^{\mathrm{T}}\cdot\overline{\boldsymbol{\omega}}+b
=\mathbf{K}_*^{\mathrm{T}}\cdot\overline{\boldsymbol{\varphi}}+b,
\label{eq:accuracy-based-improvement-024} 
\end{equation}
depends on a set of support vectors which are the data points most
difficult to regress and have direct bearing on the optimum location
of the target PES, as illustrated by Figure \ref{fig:svr-prin-00}.
Therefore, the support vector lies on and outside
the two hypersurfaces that are parallel to the target prediction
$y(\mathbf{x})$ with the distance of given tolerance error $\pm\epsilon$.
In Equation \eqref{eq:accuracy-based-improvement-024},
$\overline{\boldsymbol{\omega}}$ and $\overline{\boldsymbol{\varphi}}$
are optimized parameter matrices and
$\mathbf{K}_*^{\mathrm{T}}=\boldsymbol{\phi}^{\mathrm{T}}\boldsymbol{\Phi}$
is $1\times p$ kernel-function matrix, where $\boldsymbol{\phi}$ and
$\boldsymbol{\Phi}$ are feature space of $\mathbf{x}$ and $\mathbf{X}$,
respectively, while $p$ is dimensionality of the feature space
\cite{son22:1983}. We refer the reader to Reference \cite{son22:1983}
for numerical details on SVR.

Similar to CPD-GPR \cite{son22:11128}, according to Equation \eqref{eq:accuracy-based-improvement-024},
the SVR potential function $V^{(\mathrm{SVR})}(\mathbf{x})=y(\mathbf{x})$
can be expanded as
\begin{equation}
V_I^{(\mathrm{SVR})}=V^{(\mathrm{SVR})}(\mathbf{x}_I)=
V^{(\mathrm{SVR})}_{i_1,\cdots,i_f}=\sum_{r=1}^{n_{\mathrm{sv}}}
\overline{\varphi}_r
K\big(\mathbf{x}_I,\mathbf{X}_r\big)+b,
\label{eq:cpd-gpr-001}
\end{equation}
where $K(\mathbf{x}_I,\mathbf{X}_r)$ and $\overline{\varphi}_r$ are the
$r$-th elements of the $\mathbf{K}_*^{\mathrm{T}}$ and $\overline{\boldsymbol{\varphi}}$
matrices, respectively, which $\mathbf{x}_I$ is coordinate matrix with
elements $\{q_{i_{\kappa}}^{(\kappa)}\}_{\kappa=1}^f$. We further assume
that the multi-dimension kernel function $K(\cdot,\cdot)$ is given in
a product form of $\{K^{(\kappa)}(\cdot,\cdot)\}_{\kappa=1}^f$ that
depend on only one variable,
\begin{equation}
K\big(\mathbf{x},\mathbf{X}_r\big)=\prod_{\kappa=1}^fK^{(\kappa)}
\big(q^{(\kappa)},Q^{(\kappa)}_r\big),
\label{eq:cpd-gpr-002}
\end{equation}
where $Q^{(\kappa)}_r$ is the $\kappa$-th element of the coordinate
matrix $\mathbf{X}_r$. Here, we would like to remind that $\mathbf{X}_r$
is the $r$-th coordinates set of the training data and thus it is already
known. Substituting Equation \eqref{eq:cpd-gpr-002} into Equation
\eqref{eq:cpd-gpr-001}, we finally obtain the CPD-SVR potential values
on grids,
\begin{equation}
V_I^{(\mathrm{CPD-SVR})}=\sum_{r=1}^{n_{\mathrm{sv}}}\overline{\varphi}_r
K\big(\mathbf{x}_I,\mathbf{X}_r\big)+b
=\sum_{r=1}^{n_{\mathrm{sv}}}\overline{\varphi}_r\prod_{\kappa=1}^f
K^{(\kappa)}\big(q^{(\kappa)}_{i_{\kappa}},Q^{(\kappa)}_r\big)+b
=\sum_{r=1}^{n_{\mathrm{sv}}}\overline{\varphi}_r
\prod_{\kappa=1}^fK^{(\kappa)}_{r,i_{\kappa}}+b,
\label{eq:cpd-gpr-003}
\end{equation}
where $K^{(\kappa)}_{r,i_{\kappa}}=K^{(\kappa)}(q^{(\kappa)}_{i_{\kappa}},Q^{(\kappa)}_r)$
means value of the kernel function for the $\kappa$-th DOF on the
$i_{\kappa}$-th grid. Comparing Equation \eqref{eq:cpd-gpr-003} with
Equation \eqref{eq:cpd-potential-form-02}, the potential tensor
$V_I^{(\mathrm{CPD-SVR})}$ already has the CPD form. Only restriction
for such expression is the assumption of Equation \eqref{eq:cpd-gpr-002}.
Using the denotation in Equation \eqref{eq:cpd-gpr-003}, the kernel
function in this work for the $\kappa$-th DOF is given by
\begin{equation}
K^{(\kappa)}\big(q^{(\kappa)},Q_r^{(\kappa)}\big)
=\exp\left(-\frac{\big(q^{(\kappa)}-Q_r^{(\kappa)}\big)^2}{2l^2}\right),
\label{eq:kernel-function}
\end{equation}
where $l$ is hyper-parameter. As already expected, since $n_{\mathrm{sv}}<n$
the CPD rank by CPD-SVR is smaller than that by CPD-GPR. The present
CPD-SVR approach is thus able to obtain lower-rank CPD form.

At the first glance, the present CPD-SVR approach is somewhat trivial
from the viewpoint of mathematics due to the simple revisions on potential
model and the kernel function. Comparing CPD-GPR \cite{son22:11128},
the present CPD-SVR approach employs sparser inner-product space as
shown by Equation \eqref{eq:cpd-gpr-003}. However, noting that SVR
is just a mathematical tool in fitting the PES, it must be further
revised and implemented to make it possible in application scenario.
This means that a special algorithm has to be designed to implement
our own goal, that is the CPD construction with low rank. We would
like to emphasize that improvements on the PES construction might not
have only a single goal but also be applicable to subsequent quantum
dynamics calculations. The present goal does not only focus on the
fitting error but also does focus on reducing the CPD rank. Furthermore,
the SVR method consumes lots of computational cost in selecting the
support vectors because of the sequential minimal optimization (SMO)
algorithm \cite{son22:1983}. Therefore, let us turn to the warm-started
algorithm for accelerating the present CPD-SVR approach.

\subsection{Warm-Started SVR Algorithm\label{sec:svr-pes-app-1}}

The SVR calculation requires that the training energy data are added
into the optimization procedure in batches or one-by-one and hence
consumes lots of computational time. To overcome this problem, a
warm-started SVR scheme, denoted by ws-SVR, is proposed in this work.
This approach needs an existing potential function $U(\mathbf{x})$
which might be crudely constructed by either regression or interpolation
through the some training database that will be used in either ws-SVR
or CPD-ws-SVR calculations. Moreover, function $U(\mathbf{x})$ must be
cheap but should correctly predict the PES topology. We define difference
function between the ws-SVR potential and the crude potential in the
form
\begin{equation}
F(\mathbf{x})=\mathbf{K}_*^{\mathrm{T}}\cdot\boldsymbol{\varphi}+b+U(\mathbf{x}).
\label{eq:warm-started-00-svr}
\end{equation}
With function $F(\mathbf{x})$ of Equation \eqref{eq:warm-started-00-svr},
two rounds of optimization should be launched. The first round of
optimization with $\mathfrak{m}$ training data requires the error
function for the $i$-th training energy
\begin{equation}
h(\mathbf{X}_i)=F(\mathbf{X}_i)-E_i=
\sum_{j=1}^{\mathfrak{m}}\varphi_jK_{ij}+b+U(\mathbf{X}_i)-E_i,
\quad i=1,2,\cdots,\mathfrak{m},
\label{eq:warm-started-01-svr}
\end{equation}
where $K_{ij}=K(\mathbf{X}_i,\mathbf{X}_j)$. Turning to $\mathfrak{m}$
training data, one may sample these data by an appropriate and
chemistry-informed distribution in the configurational space making the
resulting error as small as possible. However, the present purpose is
reduction of the CPD rank by SVR, instead of sampling of these initial
data. For simple, the present $\mathfrak{m}$ training data are randomly
distributed to show the implementation of SVR. Next, setting tolerance
error $\epsilon>0$, the $i$-th training point $\{\mathbf{X}_i,E_i\}$
must belong to one of the following three sets. First, if
$h(\mathbf{X}_i)\in(-\epsilon,+\epsilon)$ then $\{\mathbf{X}_i,E_i\}$
belongs to remain set $\mathcal{R}$ where corresponding coefficient
$\varphi_i$ is equal to zero. Second, if $h(\mathbf{X}_i)=\pm\epsilon$
then $\{\mathbf{X}_i,E_i\}$ belongs to support set $\mathcal{S}$ with
$\varphi_i\in(-c,+c)$. Finally, if $h(\mathbf{X}_i)\in(-\infty,-\epsilon)\cup(+\epsilon,+\infty)$
then $\{\mathbf{X}_i,E_i\}$ belongs to error set $\mathcal{E}$ with
$\varphi_i=\pm c$. Here, $c$ is a given hyper-parameter of the regression
model. With the above definitions, one can determine classifications of
a total of $\mathfrak{m}$ training points after this round of optimization.

Now, the second round of optimization could be performed by adding one
new point $\{\mathbf{X}_{\ell},E_{\ell}\}$ in the training database
and then new classifications of these $\mathfrak{m}+1$ training points
have to be determined. After the second round of optimization, it is
easy to find that $F(\mathbf{X}_i)$ becomes
\begin{equation}
F(\mathbf{X}_i)=\sum_{j=1}^{\mathfrak{m}}\varphi_j'K_{ij}+
\varphi_{\ell}K_{i\ell}+b'+U(\mathbf{X}_i),
\label{eq:warm-started-02-svr}
\end{equation}
and hence the error function becomes
\begin{equation}
h(\mathbf{X}_i)=\sum_{j=1}^{\mathfrak{m}}\varphi_j'K_{ij}+
\varphi_{\ell}K_{i\ell}+b'+U(\mathbf{X}_i)-E_i.
\label{eq:warm-started-03-svr}
\end{equation}
The apostrophe indicates that the corresponding parameter is updated
due to a second round of optimization. Comparing Equation
\eqref{eq:warm-started-03-svr} with Equation \eqref{eq:warm-started-01-svr}
the difference between two optimizations by adding the $\ell$-th
training point is given by
\begin{equation}
\Delta h(\mathbf{X}_i)=
\sum_{j=1}^{\mathfrak{m}}\big(\varphi_j'-\varphi_j\big)K_{ij}
+\varphi_{\ell}K_{i\ell}+(b'-b).
\label{eq:warm-started-04-svr}
\end{equation}
By Equation \eqref{eq:warm-started-04-svr}, if the KKT condition is met
and all of training points always belong to the original set, only those
in $\mathcal{S}$ are changed making $h(\mathbf{X}_k)=\pm\epsilon$ a
constant and $\Delta h(\mathbf{X}_k)=0$, implying the expression
\begin{equation}
\sum_{s\in\mathcal{S}}\big(\varphi_s'-\varphi_s\big)K_{is}+
\varphi_{\ell}K_{i\ell}+(b'-b)
=\sum_{s\in\mathcal{S}}\Delta\varphi_sK_{is}+\varphi_{\ell}K_{i\ell}+\Delta b
=0.
\label{eq:warm-started-05-svr}
\end{equation}
By Equation \eqref{eq:warm-started-05-svr}, it is possible to determine
new classification for the $i$-th training point $\{\mathbf{X}_i,E_i\}$
leading to the working equations of wa-SVR. Derivations by Equations
\eqref{eq:warm-started-04-svr} and \eqref{eq:warm-started-05-svr} are
given in the Supporting Information file. We only give a brief description
of the working equations themselves.

\subsection{Selecting Process of Support Vectors\label{sec:select-sv}}

By the KKT condition, we have $\sum_{i=1}^{\mathfrak{m}}\varphi_i=0$
and hence $\sum_{i=1}^{\mathfrak{m}}(\varphi_i'-\varphi_i)+\varphi_{\ell}=0$.
Since only the training point in $\mathcal{S}$ is able to change the
coefficients, in Supporting Information we prove that
\begin{align}
\sum_{s\in\mathcal{S}}\big(\varphi_s'-\varphi_s\big)+\varphi_{\ell}&=0,
\label{eq:warm-started-06-svr}
\allowdisplaybreaks[4] \\
\left(\begin{array}{c}
\Delta b \\
\Delta\varphi_{s_1}  \\
\cdots  \\
\Delta\varphi_{s_\mathfrak{m}}
\end{array}\right)
=-\left(\begin{array}{cccc}
0 & 1 & \cdots & 1 \\
1 & K_{s_1s_1} & \cdots & K_{s_1s_\mathfrak{m}}  \\
\cdots & \cdots & \cdots & \cdots \\
1 & K_{s_\mathfrak{m}s_1} & \cdots & K_{s_\mathfrak{m}s_\mathfrak{m}}  \\
\end{array}\right)^{-1}
\left(\begin{array}{c}
1 \\
K_{s_1\ell} \\
\cdots  \\
K_{s_\mathfrak{m}\ell}
\end{array}\right)\varphi_{\ell}
&=\boldsymbol{\lambda}\varphi_{\ell},
\label{eq:warm-started-11-svr}
\end{align}
where $\{s_i\}_{i=1}^{\mathfrak{m}}$ is set of indices for the training
points in $\mathcal{S}$, $\boldsymbol{\lambda}$ a $(\mathfrak{m}+1)\times1$
matrix with $\lambda_j$ as elements. By Equation \eqref{eq:warm-started-11-svr},
it is possible to update $\boldsymbol{\varphi}$ and $b$ leading to its
name of updating equation. Now, let us show how to update function $h(\mathbf{X}_i)$.
Since the error funciton of support vectors are always equal to $\pm\epsilon$,
only the training data in $\mathcal{R}$ or $\mathcal{E}$ change the error
function. Thus, by indices $\{e_i\}_{i=1}^{\mathfrak{m}}$ for data in
$\mathcal{R}$ or $\mathcal{E}$, Equation \eqref{eq:warm-started-04-svr}
can be rewritten in the matrix form
\begin{equation}
\left(\begin{array}{c}
0 \\
\Delta h(\mathbf{X}_{e_1})  \\
\cdots  \\
\Delta h(\mathbf{X}_{e_{\mathfrak{m}}})  \\
\end{array}\right)=
\left[\left(\begin{array}{c}
1  \\
K_{e_1\ell} \\
\cdots  \\
K_{e_\mathfrak{m}\ell}
\end{array}\right)+
\left(\begin{array}{cccc}
0 & 1 & \cdots & 1 \\
1 & K_{e_1s_1} & \cdots & K_{e_1s_\mathfrak{m}}  \\
\cdots & \cdots & \cdots & \cdots \\
1 & K_{e_\mathfrak{m}s_1} & \cdots & K_{e_\mathfrak{m}s_\mathfrak{m}}  \\
\end{array}\right)
\boldsymbol{\lambda}\right]\cdot
\varphi_{\ell}
=\boldsymbol{\gamma}\cdot\varphi_{\ell},
\label{eq:warm-started-13-svr}
\end{equation}
where $\boldsymbol{\gamma}$ is a $(\mathfrak{m}+1)\times1$ matrix with
$\gamma_j$ as elements. By resolving Equation \eqref{eq:warm-started-13-svr}
one can update the error funciton $h(\mathbf{X}_i)$. Equations
\eqref{eq:warm-started-11-svr} and \eqref{eq:warm-started-13-svr} are
thus working equations of the present ws-SVR approach.

\begin{figure}[h!]
  \centering
   \includegraphics[width=18cm]{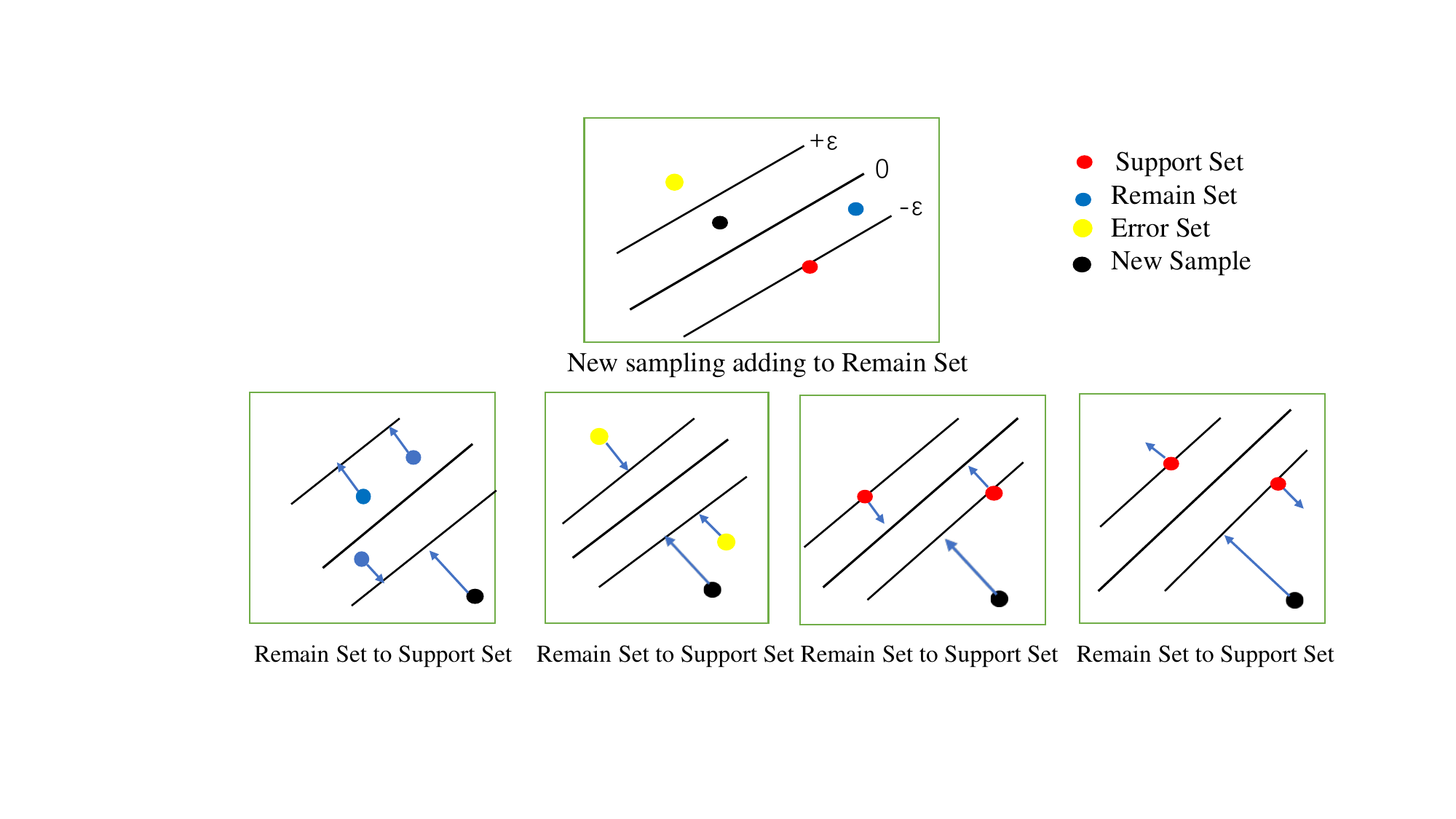}
    \caption{\figfoot}
     \label{fig:alg}
      \end{figure}

Now, We should explore the updating scheme of the classifications
of the training data and show the role of existed potential funciton
$U(\mathbf{x})$. We graphically collect updating scheme of the training
data in Figure \ref{fig:alg} and briefly describe cases of this scheme
in Table \ref{tab:alg-ws-svr}. In general, two cases for adding new data
exist in implementing ws-SVR. In Case \#1, the new point (denoted by index
$\ell$) is added into $\mathcal{R}$ implying that $\vert h(\mathbf{X}_{\ell})\vert<\epsilon$
(see the upper panel of Figure \ref{fig:alg}) and the other points are
unchanged. In Case \#2, the new point is added into $\mathcal{E}$ or
$\mathcal{S}$ making the decision surface of the last round changed
and the last classifications might be changed. However, in order to
minimize changes in one optimization round, we should keep the training
data belonging to their original sets as much as possible. This is
because the updating scheme by Equations \eqref{eq:warm-started-11-svr}
and \eqref{eq:warm-started-13-svr} requires a prerequisite that every
training data will not leave their original set implying that the
classification of the training points must be fixed. This is, of course,
not the case when a new round of optimization is finished. Careful
analysis easily indicates four different situations in changing the
classification during optimization, as given in the lower panel of
Figure \ref{fig:alg}. 

In the first situation, the $i$-th
training point is moved from $\mathcal{E}$ to $\mathcal{S}$. Since
function $h(\mathbf{X}_i)$ is changed from
$h(\mathbf{X}_i)\in(-\infty,-\epsilon)\cup(+\epsilon,+\infty)$
to $h(\mathbf{X}_i)=\pm\epsilon$, {\it i.e.}
$\Delta h(\mathbf{X}_{e_i})=\pm\epsilon-h(\mathbf{X}_{e_i})$,
Equation \eqref{eq:warm-started-13-svr} implies
$\Delta\varphi_{\ell}=\gamma_{e_i}^{-1}(\pm\epsilon-h(\mathbf{X}_{e_i}))$.
The seond situation is that the $i$-th point is moved from $\mathcal{R}$
to $\mathcal{S}$ and thus difference of the error function is
$\Delta h(\mathbf{X}_{r_i})=\pm\epsilon-h(\mathbf{X}_{r_i})$, where
$\{r_i\}_{i=1}^{\mathfrak{m}}$ represent set of indices of the training
point in $\mathcal{R}$. By updating equation (that is Equation
\eqref{eq:warm-started-11-svr}), one can have
$\Delta\varphi_{\ell}=\gamma_{r_i}^{-1}(\pm\epsilon-h(\mathbf{X}_{r_i}))$.
In the third situation, the $i$-th point is moved from $\mathcal{S}$
to $\mathcal{R}$. One can similarly have got
$\Delta\varphi_{\ell}=\lambda_{s_i}^{-1}\varphi_{s_i}$. In the fourth
situation the $i$-th point is moved from $\mathcal{S}$ to $\mathcal{E}$
implying that $\Delta\varphi_{\ell}=\lambda_{s_i}^{-1}(\pm c-\varphi_{s_i})$.
All of the above cases are collected in Table \ref{tab:alg-ws-svr}.
These cases imply that the pre-existed potential function $U(\mathbf{x})$
is helpful in updating classifications of the training points and
accelerating the SVR calculations. Due to this ability of function
$U(\mathbf{x})$, the SVR calculations become to be warm-started leading
to the name of ws-SVR. Combination ws-SVR and CPD-SVR, we finally
obtain the CPD-ws-SVR approach. Having the CPD-ws-SVR method, we
should inspect its power in the PES construction. Collected in Table
\ref{tab:examples} are three examples for testing the CPD-ws-SVR
approach, in particular for testing the ws-SVR implementation. In
Table \ref{tab:examples}, we also give numerical details of the
present SVR calculations. These examples are H + H$_2$, H$_2$ + H$_2$,
and H$_2$/Cu(111).

\section{Results and Discussions\label{sec:results}}
\subsection{Calculations on H + H$_2$\label{sec:benchmark}}

The first benchmark is the quantum dynamics of the H + H$_2$ reaction
which is typical and simple for testing the PES. The Jacobi coordinates
set is used to define the positions of three atoms, as given in Table
\ref{tab:examples}. To define this system, the center-of-mass (COM) of
the H$_2$ molecule is denoted by G. The Jacobi coordinates set for H +
H$_2$ contains bond length $r_{\mathrm v}$ of H$_2$, dissociation length
$r_{\mathrm d}$ between the impacting H atom and G, and their included
angle $\theta$. With the Jacobi coordinates, the Hamiltonian operator is
given by \cite{son19:114302}
\begin{equation}
\hat{H}_{\mathrm{H}_3}=
-\frac{1}{2M}\frac{\partial^2}{\partial r^2_{\mathrm d}}-
\frac{1}{2\mu}\frac{\partial^2}{\partial r^2_{\mathrm v}}
-\left(\frac{1}{2Mr^2_{\mathrm d}}+\frac{1}{2\mu r^2_{\mathrm v}}\right)
\left(\frac{1}{\sin\theta}\frac{\partial}{\partial\theta}\sin\theta
\frac{\partial}{\partial\theta}\right)+
V_{\mathrm{H}_3}(r_{\mathrm d},r_{\mathrm v},\theta),
\label{eq:h3-ham}
\end{equation}
where $M$ and $m$ are reduced masses along $r_{\mathrm d}$ and
$r_{\mathrm v}$, respectively. In Equation \eqref{eq:h3-ham},
the PES term $V_{\mathrm{H}_3}(r_{\mathrm d},r_{\mathrm v},\theta)$
in CPD form are constructed through the present CPD-ws-SVR approach
on the basis of the training data prepared by pre-existed BKMP2 PES,
where the potential energy of thoroughly separated three H atoms is
set to be zero. Unless otherwise specified, the atomic units are used
in this paper. Based on the Hamiltonian operator given by Equation
\eqref{eq:h3-ham}, quantum dynamics calculations are performed using
the Heidelberg MCTDH package \cite{mctdh:MLpackage}. We refer the
reader to References \cite{mey90:73,man92:3199,bec00:1} for details
of MCTDH. In the present MCTDH calculations, the vibration ground
state of the H$_2$ molecule is used as reactants, while the third H
atom is located at $r^{(0)}_{\mathrm d}=4.5$ bohr with an initial
momentum of $8.0$ au moving towards H$_2$. Here, the ranges of
coordinates are $1.00\leq r_{\mathrm d}\leq 9.04$, $0.60\leq r_{\mathrm v}\leq 6.24$,
and $0\leq\theta\leq\pi/2$, while $68$, $48$, and $31$ grid points are
used, respectively. To ensure convergence in MCTDH calculations, the
single-particle function (SPF) size of $r_{\mathrm d}/r_{\mathrm v}/\theta=20/20/20$
is used. Numerical details of the present dynamics calculations are
collected in Table \ref{tab:numerical-details}.

\begin{figure}[h!]
  \centering
   \subfigure[\quad Dependance of tolerance and error]{%
    \includegraphics[width=16cm]{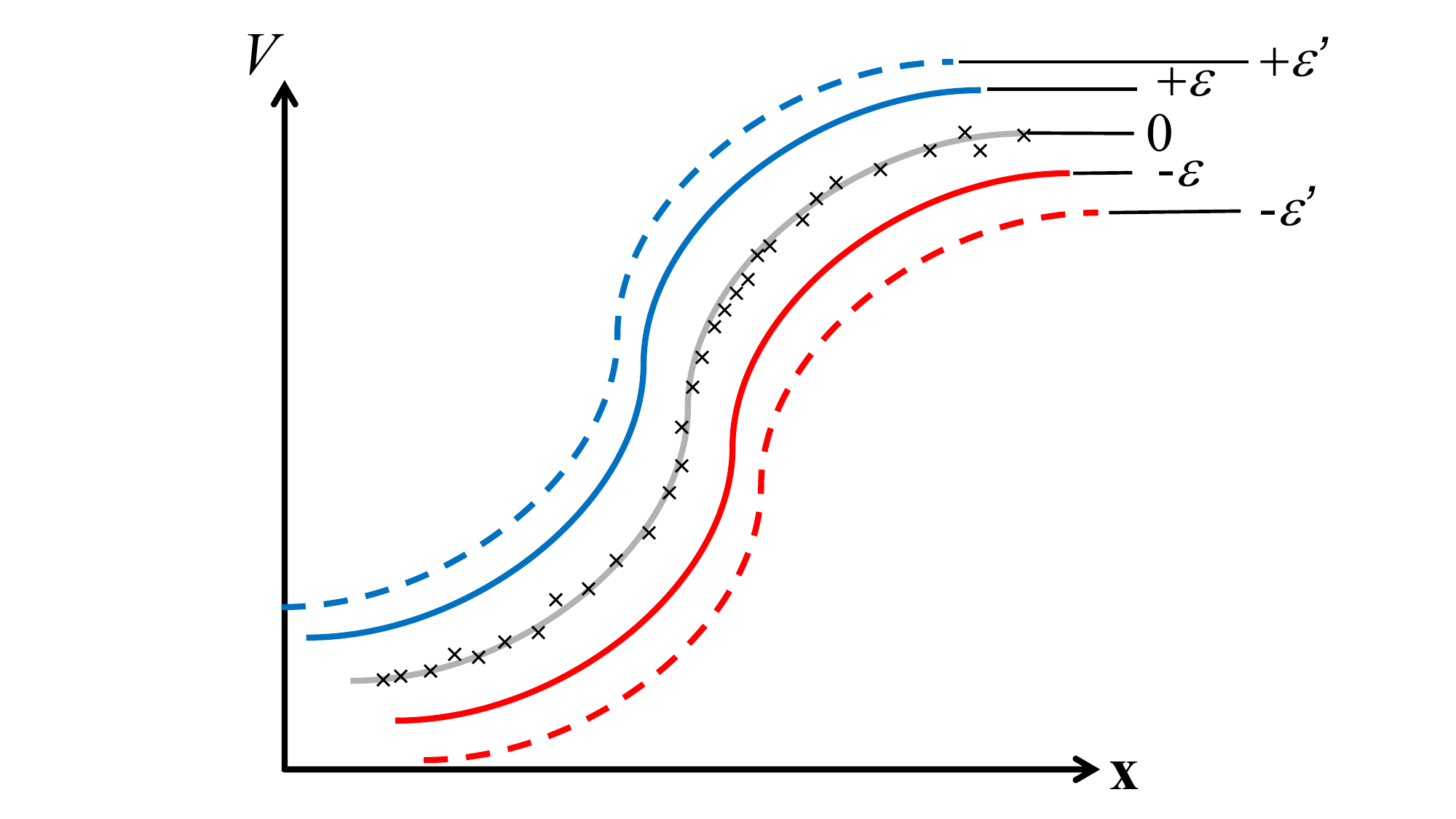}
     }
      \caption{\figfoot}
       \label{fig:svr-prin-00}
        \end{figure}

Turning to CPD-ws-SVR, the one-dimensional kernel functions for all
DOFs have the same form, as given by Equation \eqref{eq:kernel-function}.
To test the resulting PES, the validation data are randomly selected
in the whole configuration space. The fitting errors of the CPD-ws-SVR
calculations are illustrated in Figure \ref{fig:conver} as function of
number of training data $n$. These errors clearly indicate the convergence
of the size of database. For instance, setting $\epsilon=10^{-4}$ hartree
the present CPD-ws-SVR calculations predict convergence at $n>7000$
where the root mean squared error (RMSE) approaches to $2\times10^{-3}$
hartree ({\it i.e.}, $<54$ meV). Moreover, as illustrated in Figure
\ref{fig:conver}, the convergence curves for RMSE as function of $n$
weakly depend on the tolerance error $\epsilon$. This feature is clear
for the blue and black curves of Figure \ref{fig:conver} where $\epsilon$
is $10^{-3}$ and $10^{-4}$ hartree. Generally, the smaller the tolerance
error is, the smaller the RMSE value should be but the larger the
computational cost should be. This implies the dependance of the
convergence curves on $\epsilon$. As previously discussed \cite{son22:1983},
however, since no noise exists in the training database, if the tolerance
error is small enough the fitting RMSE depends on $\epsilon$ weakly,
or even independs on $\epsilon$. This can be understood with the aid
of Figure \ref{fig:svr-prin-00}(a). The support vectors lie on and
outside the two hypersurfaces that are parallel to the target prediction
function with distances of tolerance error $\pm\epsilon$. The optimal
potential function stems from the function class with the biggest
capacity of independent samples that one can twiddle keeping the
functional form not too complex. If no noise exists in the training
database, the support vectors should independ on the choice of $\epsilon$.
This point is clear by comparing the solid lines with dashed lines in
Figure \ref{fig:svr-prin-00}(b). Therefore, cost-performance of SVR
might become lower when $\epsilon$ is too small.

Furthermore, Figure \ref{fig:supp-vec} illustrates distributions of
support vectors associated with the potential energy, including
distributions of $n_{\mathrm{sv}}/n$ and $n_{\mathrm{sv}}$. It is clear
that the ratio $n_{\mathrm{sv}}/n$ approaches to roughly $80\%$ at
lower-lying energy region, while such ratio slowly approaches to about
$60\%$ at middle-lying energy region and then quickly decreases to
about $30\%$ at higher-lying energy region. Because of Equation
\eqref{eq:cpd-gpr-003}, comparing with the CPD-GPR approach \cite{son22:11128},
the CPD rank computed by CPD-ws-SVR is reduced by a factor of $\sim70\%$
making the expansion expression of Equation \eqref{eq:cpd-gpr-003} more
contracted than that of CPD-GPR. The fact that the ratio $n_{\mathrm{sv}}/n$
is a decreasing function of potential energy, as given by Figure
\ref{fig:supp-vec}, might be caused by different importance of different
energy regions. Since the lower-lying region plays much more important
role in chemistry dynamics than the middle- and higher-lying regions,
more attention must be paid on lower-lying region in regression. In
lower-lying energy regions which contain reactants, products, transition
state, and probable intermediate, the reaction occurs along these steady
points. In other region with higher-lying potential energy, the potential
function is either flat if distances among atoms are very large or steeply
increases if one or some of these distances are very small. Because of
existences of the steady points, the first- or second-order derivatives
in lower-lying energy regions usually vary more significantly than those
in higher-lying energy regions, say those with large distances among the
atoms. Since the lower-lying energy regions contain all of steady points
along the reaction coordinate and hence all of information in the reaction,
the learning processes for lower-lying energy regions are more complex
than those for higher-lying energy regions. Such region is hence difficult
to build making the ratio $n_{\mathrm{sv}}/n$ larger than that of other
region.

\begin{figure}[h!]
  \centering
   \includegraphics[width=16cm]{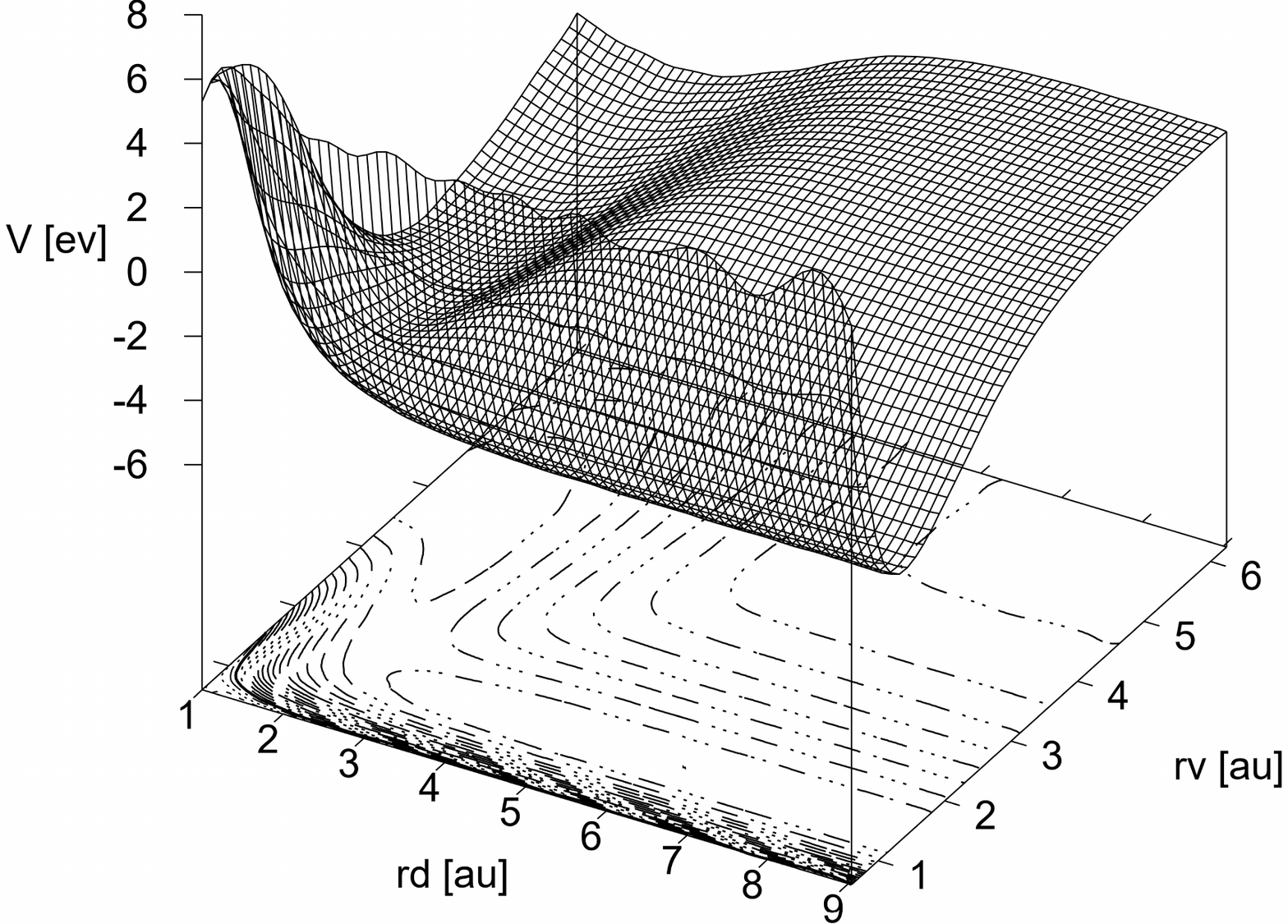}
    \caption{\figfoot}
     \label{fig:pes-contor}
      \end{figure}

\begin{figure}[h!]
  \centering
   \includegraphics[width=16cm]{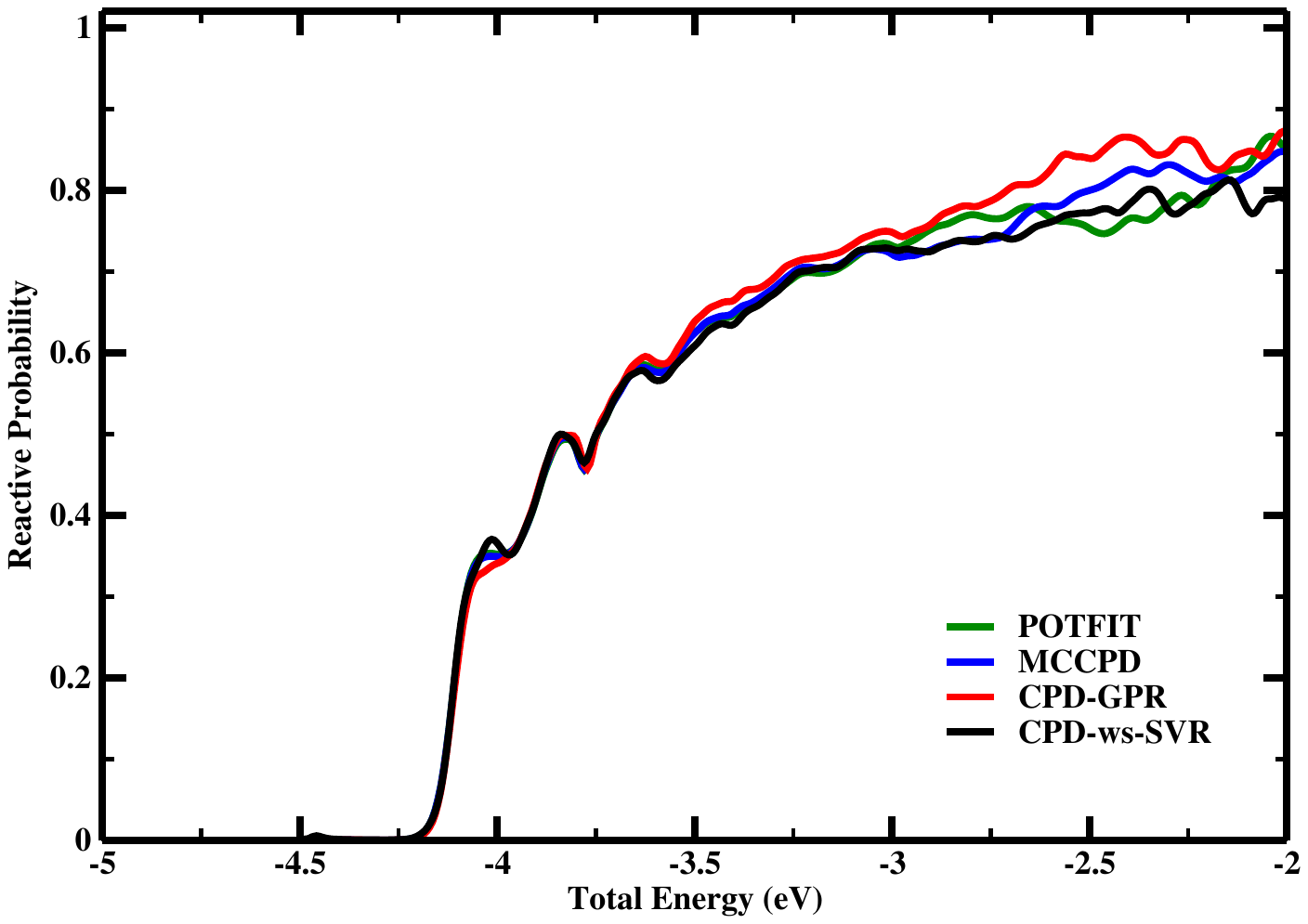}
    \caption{\figfoot}
     \label{fig:flux}
      \end{figure}

Finally, illustrated in Figure \ref{fig:pes-contor} gives two-dimension
contour cut together with three-dimension surface of the present
CPD-ws-SVR PES at $\theta=\pi/2$ where $n=9000$ and $\epsilon=10^{-4}$
hartree while the number of support vectors is determined to be
$n_{\mathrm{sv}}=5726$. By Figure \ref{fig:pes-contor} one
can clearly find that the topology of the present PES with RMSE of
$<1.8\times10^{-3}$ hartree ({\it i.e.} $<50$ meV) is satisfactory
because this RMSE value approaches to the chemical accuracy ($\sim43$ meV).
Optimzing the distribution of the training data and hyper-parameters
of the SVR model \cite{son22:1983}, the CPD-ws-SVR calculations
indicate the RMSE of $\sim10^{-4}$ hartree ($\sim2.7$ meV) which
is within the results by standard tools. To further test the present
CPD-ws-SVR calculations, extensive MCTDH calculations are performed on
the PESs that have been constructed through various methods. Illustrated
in Figure \ref{fig:flux} are present reactive probability curves comparing
with previously reported curves \cite{son22:11128}. To compute these
curves, the PESs in CPD or SOP were constructed through POTFIT, MCCPD,
and CPD-GPR, while the PES cpnstructed by SVR is shown by Figure \ref{fig:pes-contor}.
As shown in Figure \ref{fig:flux}, a good agreement among these reactive
probabilities can be found when the total energy is smaller than $-2.0$
eV. This clearly implies accuracy of the present CPD-ws-SVR calculations,
even though the distribution of training data and hyper-parameters of SVR
are not optimized.

\subsection{Calculations on H$_2$ + H$_2$ and H$_2$/Cu(111)\label{sec:otherwise}}

Having considered the H + H$_2$ system in Section \ref{sec:benchmark},
we turn to the testing calculations for the H$_2$ + H$_2$ and H$_2$/Cu(111)
systems whose coordinates are given in Table \ref{tab:examples}. To
construct their PESs, the training data are computed through existing
PESs \cite{boo02:666,dia10:6499,tho12:8628} instead of {\it ab initio}
energy calculations because the main purpose of the present calculations
is to show the power of CPD-ws-SVR. For each system, we randomly sample
the training data and validate data forming training set and validate
set, respectively, composing of a total of $5\times10^5$ energy data.
Then, 6D CPD-GPR and CPD-ws-SVR calculations are performed to build new
PESs in the CPD form. With the errors on validate sets as functions of
the number of training data $n$, one can discuss the convergence of
regressions and compare CPD-ws-SVR with CPD-GPR. Moreover, to show the
power of SVR in reducing the CPD rank, the number of support vectors
$n_{\mathrm{sv}}$ together with ratios of number of support vector to
training data $n_{\mathrm{sv}}/n$ are given as functions of the number
of training data. In the present CPD-ws-SVR calculations, the tolerance
error is set to be $10^{-4}$ hartree that is used for the case of H +
H$_2$.

Before giving numerical results of the present CPD-ws-SVR and CPD-GPR
calculations, we would like to mention that, quantum dynamics of the
H$_2$ + H$_2$ and H$_2$/Cu(111) systems based on the PESs fitted in this
work require multi-dimension MCTDH calculations. In technique, direct
MCTDH calculations by such PESs are still difficult, because such
multi-dimension MCTDH calculations need the mode combination (mc)
scheme, where two or three coordinates are combined into one logical
coordinate \cite{bec00:1}. Recently, the CPD-GPR approach has been
further implemented by us using the mode-combination scheme leading
to the CPD-mc-GPR approach \cite{son24:597}. A 9D test for OH + HO$_2$
indicates that CPD-mc-GPR has capability to build the CPD potential in
the mode-combination scheme to promote efficiency in multi-dimension
ML-MCTDH calculations \cite{son24:597}. Further testing high-dimensional
CPD-mc-GPR calculations is also planned. Similarly, to reduce the rank
of the CPD-mc-GPR potential, the SVR technique should be adopted leading
to CPD-mc-ws-SVR approach for multi-dimension MCTDH calculations. Since
many difficulties and challenges exist in implementing CPD-mc-ws-SVR,
it is planned in the future work on multi-dimension ML-MCTDH calculations
on OH + HO$_2$ $\to$ O$_2$ + H$_2$O. As will be shown in this work (see
below), the SVR technique has power to reduced the CPD rank. We believe
that further multi-dimension ML-MCTDH calculations must be possible, even
though theoretical challenges still exist as will be discussed in Section
\ref{sec:dis-num} below.

Illustrated in Figure \ref{fig:svr-con} are convergence inspections of
the present CPD-GPR and CPD-ws-SVR calculations for testing systems.
By Figure \ref{fig:svr-con}, we can easily conclude the following two
points. First, varying the number of training data the GPR calculations
are more fast approach convergence than the SVR calculations. As shown
by Figure \ref{fig:svr-con}(a), the 3D CPD-GPR calculations for H + H$_2$
approach to convergence if $n\geq5\times10^3$, while the 6D ones approach
to convergence if $n\geq1.5\times10^4$. As shown by Figure \ref{fig:conver},
however, the 3D CPD-ws-SVR calculations approach to convergence when
$n\geq8\times10^3$. Of course, convergence depends on the tolerance
error as well, as shown by Figure \ref{fig:conver}. Meanwhile, as
shown by Figure \ref{fig:svr-con}(b), the 6D CPD-ws-SVR calculations
approach to convergence when $n\geq5\times10^4$. Second, for each
system, the validate error of CPD-GPR is generally smaller than that
of CPD-ws-SVR. For H + H$_2$, the validate error of GPR is almost zero
(see also the green line of Figure \ref{fig:svr-con}(a)), but that of
CPD-ws-SVR is about $10^{-3}$ hartree (see also Figure \ref{fig:conver}).
Comparing Figure \ref{fig:svr-con}(a) with Figure \ref{fig:svr-con}(b),
the validate errors of 6D CPD-GPR are almost twice those of 6D CDP-ws-SVR.
We shall interpret these numerical results on convergence, in particular
the error performance of the present CPD-ws-SVR calculations in Section
\ref{sec:dis-num} below.

\begin{figure}[h!]
  \centering
   \includegraphics[width=16cm]{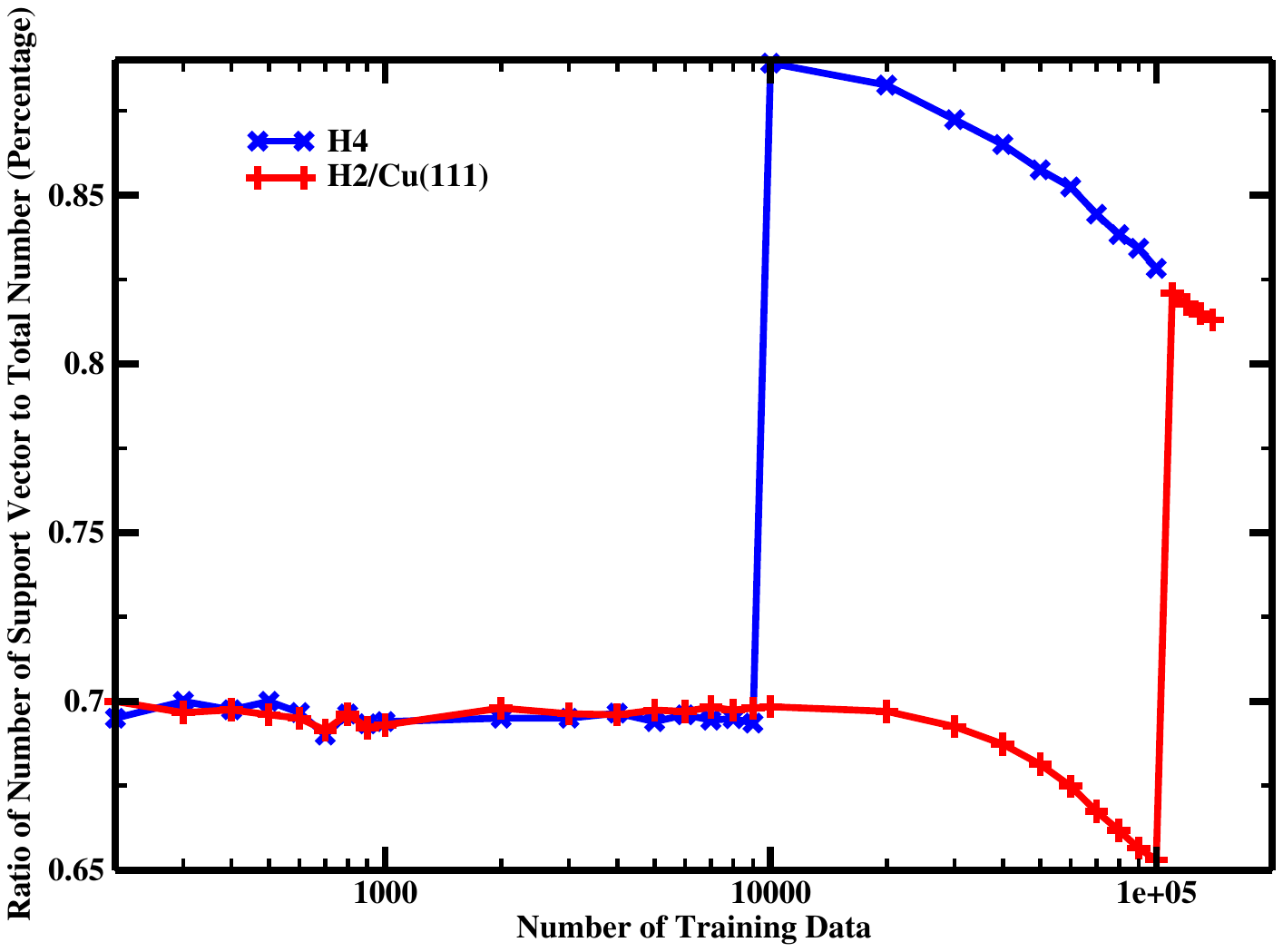}
    \caption{\figfoot}
     \label{fig:ration-sv}
      \end{figure}

Next, let us turn to the number of support vectors $n_{\mathrm{sv}}$
as function of the training data $n$. Unlike the black dots of
Figure \ref{fig:supp-vec}, given in Figure \ref{fig:ration-sv} are
rations of the number of support vectors to that of training data as
function of $n$, that is ration $n_{\mathrm{sv}}/n$ as funciton of
$n$. As usual, the blue and red lines represent results of the H$_2$
+ H$_2$ and H$_2$/Cu(111) systems, respectively. When $n\leq10^4$,
$n_{\mathrm{sv}}/n$ for H$_2$ + H$_2$ and H$_2$/Cu(111) are both
identical to roughly $0.7$ implying reduction of the CPD rank. When
$n$ increases, however, the situation becomes interesting. For the
H$_2$ + H$_2$ system (see the blue line of Figure \ref{fig:ration-sv}),
the $n_{\mathrm{sv}}/n$ curve mutates from $0.7$ to about $0.88$ at
$n\sim10^4$. Further increasing $n$, the $n_{\mathrm{sv}}/n$ curve
decreases slowly from $0.88$ to roughly $0.82$ at $n\sim10^5$. Turning
to the H$_2$/Cu(111) system, as shown by the red line of Figure
\ref{fig:ration-sv}, there exists mutatation at $n\sim10^5$. Moreover,
these two curves are approximately parallel to each other at the
interval of $10^4<n<10^5$ while they seem to become identical at
$n>10^5$. The reason why there are such features is still not clear
at the present. Exploring the thermodynamic nature of the data in
building the PES is planning.

\subsection{Discussions on Numerical Results\label{sec:dis-num}}

The convergence inspections illustrated in Figures \ref{fig:conver}
and \ref{fig:svr-con} indicate that the CPD-ws-SVR calculations
predict the resulting potential function with larger validate error
than the CPD-GPR calculations. Now, we are arriving at the position
to compare 6D CPD-ws-SVR for H$_2$ + H$_2$ and H$_2$/Cu(111) with 3D
case for H + H$_2$ and then to interpret the reasons why the present
6D CPD-ws-SVR calculations produce rather poor performance than that
for H + H$_2$. First, it should be mentioned that the present CPD-ws-SVR
calculations are performed to show the present implementation and thus
the computational setups in the SVR calculations have not yet optimized.
According to our previous discussions \cite{son22:1983}, it is still
difficult to optimize a part of hyper-parameters (see Equations (22)
and (29) of Reference \cite{son22:1983}) in optimizing the SVR target.
Since optimization algorithm of the SVR hyper-parameters is missed,
these hyper-parameters are chosen by experiences and several tests.
Due to such insufficient optimizations, the present SVR performance
is rather poor in fitting error. Second, as illustrated by Figure
\ref{fig:conver}, the tolerance error influences the validate error
of the resulting SVR potential function. Answer of the question how
to determine the tolerance error is still missed. As have discussed
in Section \ref{sec:benchmark}, for the training set without any noise,
the smaller the tolerance error is, the smaller the training error is,
even though a large amount of computational costs might be consumed.
But, the situation changes for the training set with noise. In this
case, if the tolerance error is set to be too small SVR may predict
larger validate error (see also Figure \ref{fig:svr-prin-00}).
Therefore, implementation of SVR must be much more carefully designed
to approach small enough validate error. Third, optimization targets
of GPR and SVR are different \cite{son22:1983} making the present 6D
CPD-ws-SVR calculations for H$_2$ + H$_2$ and H$_2$/Cu(111) perform
rather poor than CPD-GPR. As previously discussed \cite{son22:1983},
the GPR approach optimizes likelihood while the SVR approach, or more
precisely support vector machine (SVM), is a decision process instead
of an optimization process. Although the decision process is essentially
a kind of optimization too, the target in SVR is to select the support
vectors among the training data. Thus, one has to carefully perform the
SVR calculations for the targets to minimize fitting error. Finally, in
selecting process of SVR, the present CPD-ws-SVR method requires an
existing function $U(\mathbf{x})$ to help this process, as shown by
Figure \ref{fig:alg}. Because the existing function $U(\mathbf{x})$ is
roughly constructed and imprecise, this $U(\mathbf{x})$-aided selecting
process should be iterative to optimize number and distribution of the
support vectors. Balancing the computational cost in SVR and fitting
error, one must stop the iterative process if it approaches some
criterion.

In general, fitting error depends on various factors in practical PES
construction, such as quality and distribution of the training database
\cite{bow11:8094,li12:041103,jia13:054112,li13:204103,beh15:1032}.
The construction approach, instead of a functional model itself, may
be a factor in producing the fitting error. However, methodology is
even not the most important factor. As previously discussed \cite{son22:1983},
almost all of modern construction approaches can be merged in a unified
regression model, implying that different approaches should produce
similar fitting error. On the one hand, quality of the training data
generally represent the noises of the data. The {\it ab initio} energy
calculations often produce noises depending on the regions in the
configuration space. If the noises are too large, the PES construction
might predict rather large error \cite{son19:114302,son20:134309}. The
segmented strategy \cite{men15:101102,son20:134309} was thus proposed
to reduce the noises improving the construction quality, where the
potential functions of a small region are constructed. On the other
hand, distribution of the training data might be the most important
factor in fitting the PES. If the data sampled in the regions with
fruitful chemistry-information, relationships among these data are
useful to reproduce dynamics and easily learned by the PES model
leading to small errors.
In the present work, the training data are computed by existed PESs
which are continuous and smooth functions implying the training data
without any noise. However, these data are randomly distributed in the
whole configuration space making most of chemistry informations be
watered down. This random distribution products a rather high fitting
error if no further optimization has been made. Improivng this
distribution and sampling the training data along the reaction
coordinate, the present CPD-ws-SVR approach is able to produce the
fitting errors of $10^{-4}\sim10^{-5}$ hartree (namely $10^{0}\sim10^1$
meV), which is similar to or smaller than that from popular fitting
approaches (say NN). Very recently, a general scheme for sampling the
data was proposed by us \cite{son24:597,mia24:532} where a unsupervised
data-driven technique is adopted. Tests of this new sampling scheme was
also reported \cite{mia24:532}. Furthermore, due to the complex
relationships among the training data, the functional form must be
flexible enough to well converge the regression process implying a
large amount of (hyper-)parameters that need to be optimized. The
more flexible the potential function is, the more training data are
needed consuming a large amount of computational cost. Therefore,
special algorithms should be developed to reduce computational cost.

\begin{figure}[h!]
  \centering
    \includegraphics[width=16cm]{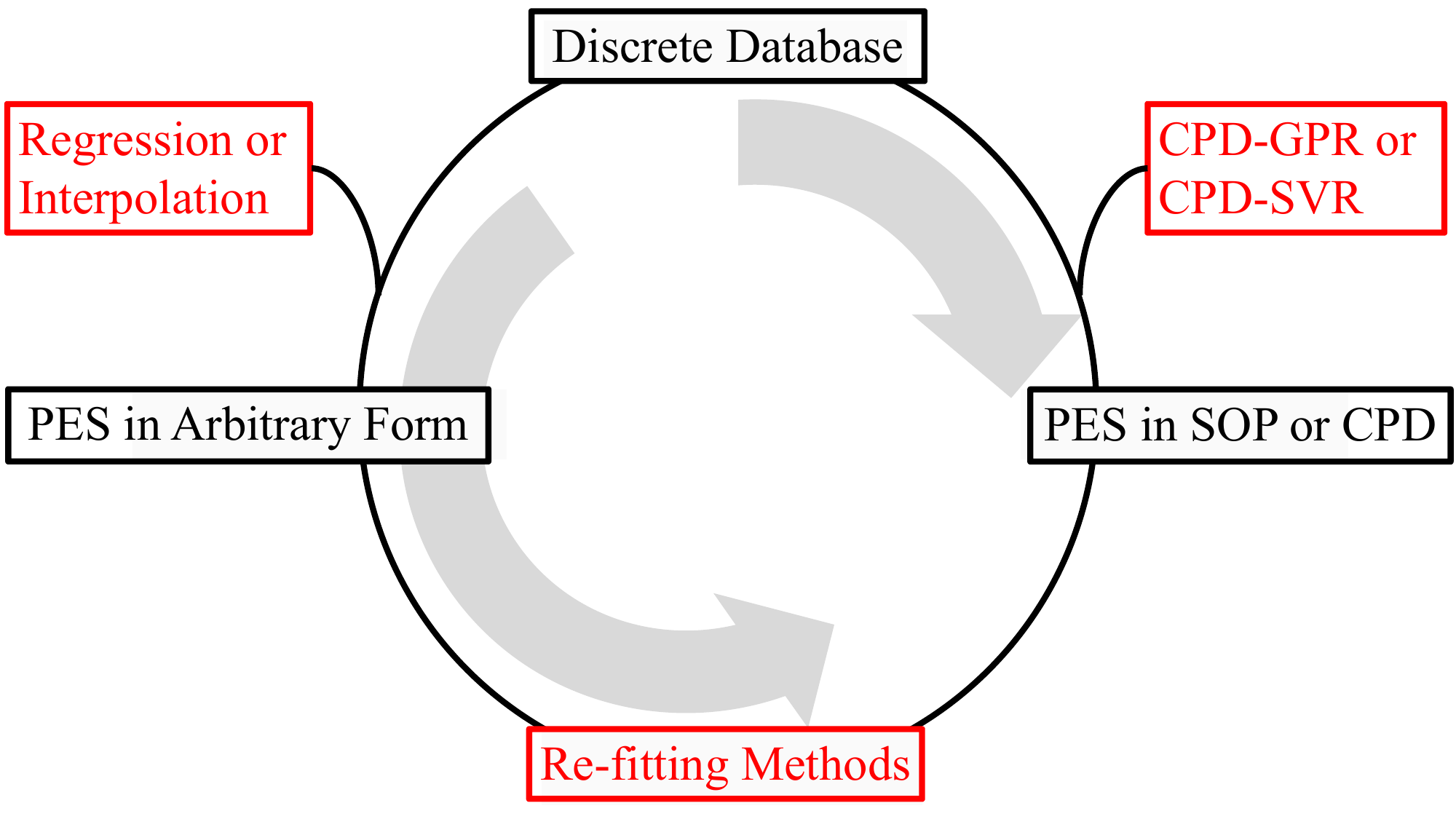}
     \caption{\figfoot}
      \label{fig:methods}
       \end{figure}

Next, turning to Figure \ref{fig:methods} let us discuss the present
one-step scheme on directly constructing the CPD potential function
and show its advantages by comparing CPD-SVR/GPR with the MCCPD method
\cite{sch20:024108,men21:2702,shi23:194102}.
First, although MCCPD is
powerful in re-fitting the PES in CPD for high-dimension system, such
as either 21D model of CO/Cu(100) \cite{men21:2702} or 75D model of
H/graphene \cite{shi23:194102}, a force-field-type potential has been
employed for the MCCPD calculations \cite{sch20:024108,men21:2702,shi23:194102}.
This is because the MCCPD approach requires a large amount of Monte
Carlo trajectories to approach convergence in integrating
\cite{sch20:024108,men21:2702,shi23:194102} and high-performance
potential is much more appropriate. In other words, either NN or
GPR PES is low-performance to implement the MCCPD calculations. This
is not surprising because high-dimension NN and GPR function is generally
very complex functional form with a large number of (hyper-)parameters
leading to lower-performance. To overcome this problem, as the second
point, we propose the one-step scheme as shown in Figure \ref{fig:methods}.
In the two-step scheme (the anticlockwise one), one has to improve the
numerical performance of the NN or GPR PES making it possible to implement
for MCCPD. However, functional flexibility and numerical performance
form a contradiction making the one-step scheme might impossible for
high-dimension systems. In this context, the two-step scheme (the
clockwise one) may be an appropriate choice. It is the present
construction scheme for the CPD potential. By CPD-GPR/SVR the
contradiction between functional flexibility and numerical performance
is totally removed. As a general rule of thumb, in fitting a PES one
should first consider its application scenario and thus one has to
carefully design the construction processes. Therefore, under the
grid-based representation one might want to choose the two-step scheme
of Figure \ref{fig:methods} in fitting the PES for quantum dynamics
calculations. Despite the above advantages, dependance of the CPD rank by
the one-step scheme, $n$ or $n_{\mathrm{sv}}$, and the question how
to further reduce the CPD rank are still open questions. One has to
overcome this disadvantage to extend the applications of the one-step
scheme.

\subsection{Discussions on Similar Methods\label{sec:dis}}

Now, other methods to construct the PES in SOP/CPD should be discussed.
Part of discussions have been simply given in Reference \cite{son22:11128}.
The first point is relationship among SOP-NN, CPD-GPR, and CPD-SVR which
belong to the one-step scheme. The SOP-NN approach \cite{man06:194105,koc14:021101}
uses a single-layer NN function to achieve a similar degree of complexity
and non-linearity as multi-layer NN in the SOP form. The SOP-NN potential
values on grids $I$ are given by \cite{man06:194105,koc14:021101}
\begin{equation}
V_I^{(\mathrm{SOP-NN})}=b_1^{(2)}+\sum_{m=1}^{\eta}w_{1m}^{(2)}
\prod_{\kappa=1}^fF^{(\kappa)}\big(b_{m\kappa}^{(1)}
+w_{m\kappa}^{(1)}q^{(\kappa)}_{i_{\kappa}}\big),
\label{eq:nn-sop-form-111}
\end{equation}
where $\{b_{m\kappa}^{(1)},b_1^{(2)}\}$ is biases set and
$\{w_{m\kappa}^{(1)},w_{1m}^{(2)}\}$ is weights set while $\eta$ is
the number of neurons. Function $F^{(\kappa)}(\cdot)$ is one-dimension
activation function. Obviously, the SOP-NN approach actually produces
the PES in the CPD form and hence it can be denoted by CPD-NN instead
of SOP-NN. Product of all one-dimension activation functions is hence
the activation function of the whole NN function.
Noting that the kernel
function of CPD-GPR has the same product characteristic of one-dimension
functions, this might indicate hidden equivalence between SOP-NN and
CPD-GPR. As previously discussed \cite{son22:1983}, this is not
surprising because the single-layer NN function is essentially equivalent
to the GPR prediction. Moreover, the SVR approach can be considered to
lie between the generalized linear regression (GLR), say NN, and kernel
model for regression (KMR), say GPR. The present CPD-SVR approach
can be considered to lie between SOP-NN and CPD-GPR and has some of
their common features. For instance, number of support vectors $n_{\mathrm{sv}}$
is smaller than that of training data $n$ and approximately equal to
that of neurons $\eta$. Moreover, SVR requires kernel function instead
of activation function making SVR similar to a kind of KMR.

Second, let us pay more attention on relationship between SOP-NN and
CPD-SVR. As given by Equations \eqref{eq:cpd-gpr-003} and
\eqref{eq:nn-sop-form-111} rank of $V_I^{(\mathrm{SOP-NN})}$ is number
of neurons $\eta$ while that of $V_I^{(\mathrm{CPD-SVR})}$ is number
of support vectors $n_{\mathrm{sv}}$. We have shown the argument of
$n_{\mathrm{sv}}\sim\eta$. In principle, CPD-SVR is able to predict
a more compact expansion form than SOP-NN, namely $n_{\mathrm{sv}}\lesssim\eta$,
due to the following several points.
First, the single-layer NN function
$V^{(\mathrm{SOP-NN})}$ needs lots of neurons to approach convergence
leading to the rank of $V_I^{(\mathrm{SOP-NN})}$ might be very high
for complex multi-dimension systems. It was proved \cite{son22:1983}
that GPR is equivalent to single-layer NN with infinite neurons and
thus the single-layer NN function predict the CPD form with very large
rank ({\it i.e.}, $\eta\to\infty$) if it approaches convergence. This
might greatly limit application of SOP-NN to multi-dimension systems.
Obviously, it is still insufficient to obtain the conclusion of
$n_{\mathrm{sv}}\lesssim\eta$ leading to the second point. It has been
proved \cite{kam18:241702,kre19:13392,var19:022001,son19:114302,son20:134309,son22:1983}
that the KMR methods, such as GPR, are easier to handle benefiting from
the Bayesian model allowing one to construct the PES with fewer training
energy data. In this context, the rank of $V_I^{(\mathrm{CPD-GPR})}$
may be not too large, namely $n\lesssim\eta$. Further noting
$n_{\mathrm{sv}}<n$ (see Figure \ref{fig:supp-vec} for example), the
CPD-SVR method has capability in building a more compact expansion CPD
form than the CPD-GPR method because the ratio $n_{\mathrm{sv}}/n$ must
be not larger than $1$. Therefore, one can have $n_{\mathrm{sv}}\lesssim\eta$.
Finally, like GPR, SVR is a hyper-parameter regression method in which
the SVR prediction model itself has already been optimized \cite{son22:1983}
on the basis of GLR. However, as shown in Sections \ref{sec:benchmark}
and \ref{sec:otherwise}, SVR is an incomplete hyper-parameter method
where part of hyper-parameters are not optimized and has to be
artifitially chosen. This is not the case of GPR where all of
hyper-parameters are optimized. This further indicates that SVR should
be more accurate if the hyper-parameters are totally optimized. This
is planned in our next work.

Third, similar to SOP-NN \cite{man06:194105,koc14:021101}, CPD-GPR
\cite{son22:11128}, and present CPD-ws-SVR method, the SOP/CP-FBR
methods proposed by Pel{\'a}ez and co-workers \cite{pan20:234110,nad23:114109}
have capability to construct the SOP/CPD form employing the two-step
scheme. In SOP/CP-FBR \cite{pan20:234110,nad23:114109}, one first
guesses an initial SOP/CPD form on a coarse grid. The hyper-parameters
of such form are optimized while the SOP/CPD form is interpolated into
a fine grid. To easily construct the initial SOP/CPD form, the SPPs are
replaced by an analytical functions, called Schmidt functions
\cite{pan20:234110}, which are expanded in a given polynomial series,
such as Chebyshev polynomial series. In SOP/CP-FBR, the potential tensors
on grid $I$ are given by \cite{pan20:234110,nad23:114109}
\begin{equation}
V_I^{(\mathrm{SOP-FBR})}=\sum_{j_1=1}^{m_1}\cdots\sum_{j_f=1}^{m_f}
C_{j_1,\cdots,j_f}\prod_{\kappa=1}^f\left(\sum_{\mu=1}^{t_{\kappa}}
c_{\mu,j_{\kappa}}^{(\kappa)}T_{\mu}\big(q_{i_{\kappa}}^{(\kappa)}\big)\right),
\quad
V_I^{(\mathrm{CP-FBR})}=\sum_{r=1}^R\lambda_r
\prod_{\kappa=1}^f\left(\sum_{\mu=1}^{t_{\kappa}}
c_{\mu,r}^{(\kappa)}T_{\mu}\big(q_{i_{\kappa}}^{(\kappa)}\big)\right),
\label{eq:sop-fbr-000}
\end{equation}
where $c_{\mu,j_{\kappa}}^{(\kappa)}$ and $c_{\mu,r}^{(\kappa)}$ are
expansion coefficients while $T_{\mu}(\cdot)$ is Chebyshev function.
The core tensor and expansion coefficients can be then optimized to
approach a given accuracy level. Equation \eqref{eq:sop-fbr-000} is
able to be used in any set of grids if the SPPs are interpolated through
\begin{equation}
v_j^{(\kappa)}\big(q_{i_{\kappa}}^{(\kappa)}\big)=
\sum_{\mu=1}^{t_{\kappa}}c_{\mu,j_{\kappa}}^{(\kappa)}
T_{\mu}\big(q_{i_{\kappa}}^{(\kappa)}\big).
\label{eq:sop-fbr-001}
\end{equation}
The denotions in Equations \eqref{eq:sop-fbr-000} and \eqref{eq:sop-fbr-001}
as well as numerical details of SOP/CP-FBR can be found in References
\cite{pan20:234110,nad23:114109}. The SOP/CP-FBR methods \cite{pan20:234110,nad23:114109}
have capability to save computational cost of the subsequent quantum
dynamics calculations due to the sparse character of the Tucker core
tensors.
Comparing Equation \eqref{eq:sop-fbr-000} with Equation \eqref{eq:cpd-gpr-003},
CP-FBR predicts similar functional form of CPD-SVR. This is not surprising
because both CP-FBR and CPD-GPR are concerned on the CPD form. Due to
its completeness, any kernel function can be expanded through a set of
Chebyshev basis functions and thus the CPD-SVR/GPR methods predict equivalent
CPD form of CP-FBR with $t_{\kappa}\to\infty$.

In addition to the grid-based representation, the so-called ``direct''
or on-the-fly dynamics methods exist. The past decade has seen a burst
of activity in the development of the direct methods merging the data-driven
techniques and wave function propagation schemes such as the on-the-fly
version of MCTDH. In these approaches \cite{ric17:4012,ric18:134116,ric22:209},
the PESs are generated in tandem with wave function propagation making
accurate on-the-fly simulations on inactivated motions possible. The
on-the-fly simulations circumvent the challenge of obtaining training
data by instead only demanding expensive {\it ab initio} calculations
when they are needed. To reduce the number of terms in the PES expansion
for on-the-fly MCTDH, Habershon and co-workers \cite{ric17:4012,ric18:134116,ric22:209}
proposed an algorithms for tensor decomposition of the KMR prediction.
To this end, a secondary function decomposition \cite{ric22:209} has
to be performed in addition to the KMR procedure. At the first step,
the potential function is represented at the grids along each DOF with
the remaining coordinates fixed. At the second step, couplings between
each pair of DOFs are then accounted by singular value decomposition
(SVD). One creats a two-dimensional grid of a pair of DOFs and then
evaluates the KMR prediction at each grid. The one-dimensional terms,
calculated previously in the first step, are then subtracted from
these grids to give a two-dimensional residue matrix which is subjected
to SVD. This decomposition procedure \cite{ric22:209} can be extended
to three-dimensional terms and similarly further. In such extentions,
an additive kernel including terms of at least three dimensions must
be used. Comparing this procedure of Habershon and co-workers
\cite{ric17:4012,ric18:134116,ric22:209} with the present CPD-ws-SVR
approach, the present CPD-SVR approach is a one-step procedure. It
is clearly different from the decomposition algorithms of Habershon
and co-workers \cite{ric17:4012,ric18:134116,ric22:209}, where an
existed function in general analytical form should be constructed
by KMR. In general, the latter two-step scheme \cite{ric17:4012,ric18:134116,ric22:209}
is similar to the traditional procedure of the Heidelberg version
of MCTDH \cite{jae95:5605,jae96:7974,jae98:3772}, as shown in Figure
\ref{fig:methods}. But, Habershon's two-step procedure \cite{ric17:4012,ric18:134116,ric22:209}
is employed to the on-the-fly quantum dynamics instead of the wave
function propagations on an existed PES.

\section{Conclusions\label{sec:con}}

In this work, a decoupled SVR approach in conjunction with a warm-started
scheme for direct canonical polyadic decomposition of a PES through a
set of discrete training energy data has been proposed to avoid the PES
in general form as intermediate in quantum dynamics calculations. The
present CPD-ws-SVR approach requires the multi-dimension kernel function
in a product of a series of one-dimension functions, similar to recently
developed CPD-GPR method \cite{son22:11128}. Due to the fact that only
a small set of support vectors play a role in SVR prediction, the present
CPD-ws-SVR approach could predict the CPD form in lower rank than CPD-GPR.
It should be noted that, the warm-started scheme needs a pre-existed
crude PES to save computational cost in determining support vectors.
To test the present CPD-ws-SVR approach, the PES of the H + H$_2$
system is constructed in the CPD form through discrete training
energy data which are are computed by existed BKMP2 PES. With the
the dynamics results computed by the decomposed PESs, a good agreement
among those by POTFIT, MCCPD, CPD-GPR, and CPD-ws-SVR can be clearly
found. Moreover, the 6D PESs in the CPD form of the H$_2$ + H$_2$ and
H$_2$/Cu(111) systems are constructed by CPD-GPR and CPD-ws-SVR methods
to show the power of SVR in reducing the CPD rank. These calculations
indicate that SVR is able to reduce the CPD rank by roughly $70\%\sim80\%$
depending on the number of training data. Finally, discussions on the
present CPD-ws-SVR and relationships among CPD-ws-SVR and previously
reported similar methods, such as SOP-NN, CPD-GPR, SOP-FBR, and CP-FBR,
have been given. It could be clear that the present CPD-ws-SVR method
should be useful in building an appropriate potential funciton for
quantum dynamics calculations, in particular for the MCTDH calculations,
and be helpful to inspire ideas for developing new tools in building
decomposed potential function in lower rank. Such decomposed PES might
be helpful in launching dynamics calculations based on single-particle
approximation (SPA).

\section*{Online Content}

The Supporting Information is available free of charge at https://pubs.acs.org/XXX/XXX.
The Supporting Information file contains:
(1) various construction schemes of the PES,
(2) geometries and coordinates of the three examples,
(3) derivations for the working equations of the ws-SVR method, and
(4) numerical details of the present quantum dynamics calculations on H + H$_2$.

\section*{Acknowledgments}

The authors gratefully acknowledges financial support by National
Natural Science Foundation of China (Grant No. 22273074) and Xi'an
Modern Chemistry Research Institute (Grant No. 204-J-2023-0691).



\clearpage
\begin{sidewaystable}
 \caption{%
Updating scheme of the present warm-started algorithm for building the
PES by SVR. The working equations of this scheme are given by Equations
\eqref{eq:warm-started-11-svr} and \eqref{eq:warm-started-13-svr}. In
general, the present implementation requires various updating cases for
the training points as well as error funciton and parameters. According
to the new points added into the model, two updating cases exist. It is
a simple and easy case if a new point is added into $\mathcal{R}$ because
nothing needs to be changed. The other case has four updating situations
of error function and parameters, according to the updating possibilities
of the training points. Illustrated in Figure \ref{fig:alg} are diagrammatic
sketch for updating the points. The second column gives items to estimate
the updating cases. These evaluation items are (1) the set into which
the new point is added, (2) illustration of the updating scheme, (3)
change of error function. The updating of points and functions are
also given. The other columns give the updating processes for Cases
\#1 and \#2. The rightmost four columns give four situations of Case
\#2. 
}%
\begin{center}
 \begin{tabular}{llllllccccccc}
\hline
No. &~~& Item  &~~& Case \#1 &~~& \multicolumn{7}{c}{Case \#2}  \\ \cline{7-13}
    &~~&       &~~&          &~~& Case \#2.1 &~~& Case \#2.2 &~~& Case \#2.3 &~~& Case \#2.4 \\
\hline 
1 && Adding point && Into $\mathcal{R}$ && \multicolumn{7}{c}{Into $\mathcal{E}$ or $\mathcal{S}$}  \\
2 && Illustration&~~& Upper panel of Figure \ref{fig:alg} && 
\multicolumn{7}{c}{Lower panel of Figure \ref{fig:alg}} \\
3 && Consequence &~~& $\vert h(\mathbf{X}_{\ell})\vert<\epsilon$ &&
\multicolumn{7}{c}{$\vert h(\mathbf{X}_{\ell})\vert\geq\epsilon$}  \\
4 && Updating points  &~~& Unchange the other points 
&& $\mathcal{E}\to\mathcal{S}$ && $\mathcal{R}\to\mathcal{S}$
&& $\mathcal{S}\to\mathcal{R}$ && $\mathcal{S}\to\mathcal{E}$ \\
5 && Updating $h(\mathbf{X}_i)$ &~~& --
&& $\Delta h(\mathbf{X}_{e_i})=\pm\epsilon-h(\mathbf{X}_{e_i})$
&& $\Delta h(\mathbf{X}_{r_i})=\pm\epsilon-h(\mathbf{X}_{r_i})$
&& -- && -- \\
6 && Updating $\varphi_{\ell}$ && --
&& $\Delta\varphi_{\ell}=\gamma_{e_i}^{-1}(\pm\epsilon-h(\mathbf{X}_{e_i}))$
&& $\Delta\varphi_{\ell}=\gamma_{r_i}^{-1}(\pm\epsilon-h(\mathbf{X}_{r_i}))$
&& $\Delta\varphi_{\ell}=\lambda_{s_i}^{-1}\varphi_{s_i}$
&& $\Delta\varphi_{\ell}=\lambda_{s_i}^{-1}(\pm c-\varphi_{s_i}$  \\
\hline
\end{tabular}
 \end{center}
  \label{tab:alg-ws-svr}
   \end{sidewaystable}
   
\clearpage
\begin{sidewaystable}
 \caption{%
Three typical examples for testing the present CPD-ws-SVR calculations
together with numerical details for SVR. The second column gives items
for estimating the calculations. These items contain (1) dimensionality,
(2) geometry, (3) coordinates set, (4) tolerance and validate errors,
(5) number of training data $n$ for final SVR calculations. Number of
supporting vectors $n_{\mathrm{sv}}$ and ratio $n_{\mathrm{sv}}/n$
depend on $n$ implying that $n_{\mathrm{sv}}$ has no determine value.
Definiations and geometries of these systems are also given in the
Supporting Information file. The third, fourth, and fifth columns give
details of H + H$_2$, H$_2$ + H$_2$, and H$_2$/Cu(111), respectively.
}
\begin{center}
 \begin{tabular}{llllccccc}
  \hline
No. &~~& Item &~~& \multicolumn{5}{c}{Systems}  \\  \cline{5-9}
    &~~&      &~~& H + H$_2$ &~~& H$_2$ + H$_2$ &~~& H$_2$/Cu(111) \\
\hline
1 && Dimensionality  && $3$ && $6$ && $6$ \\
2 && Geometry        &&
\includegraphics[width=5cm]{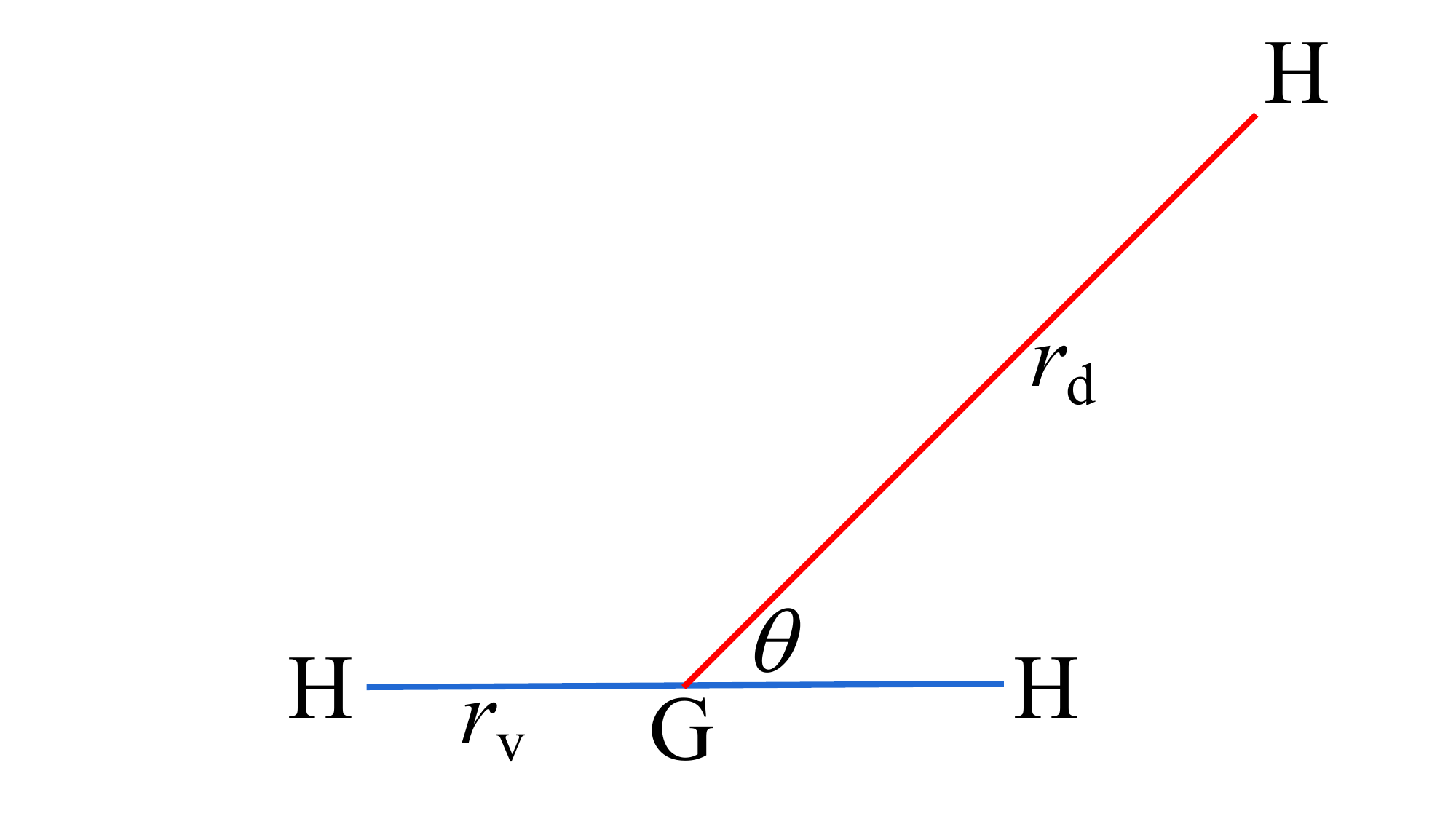} &&
\includegraphics[width=5cm]{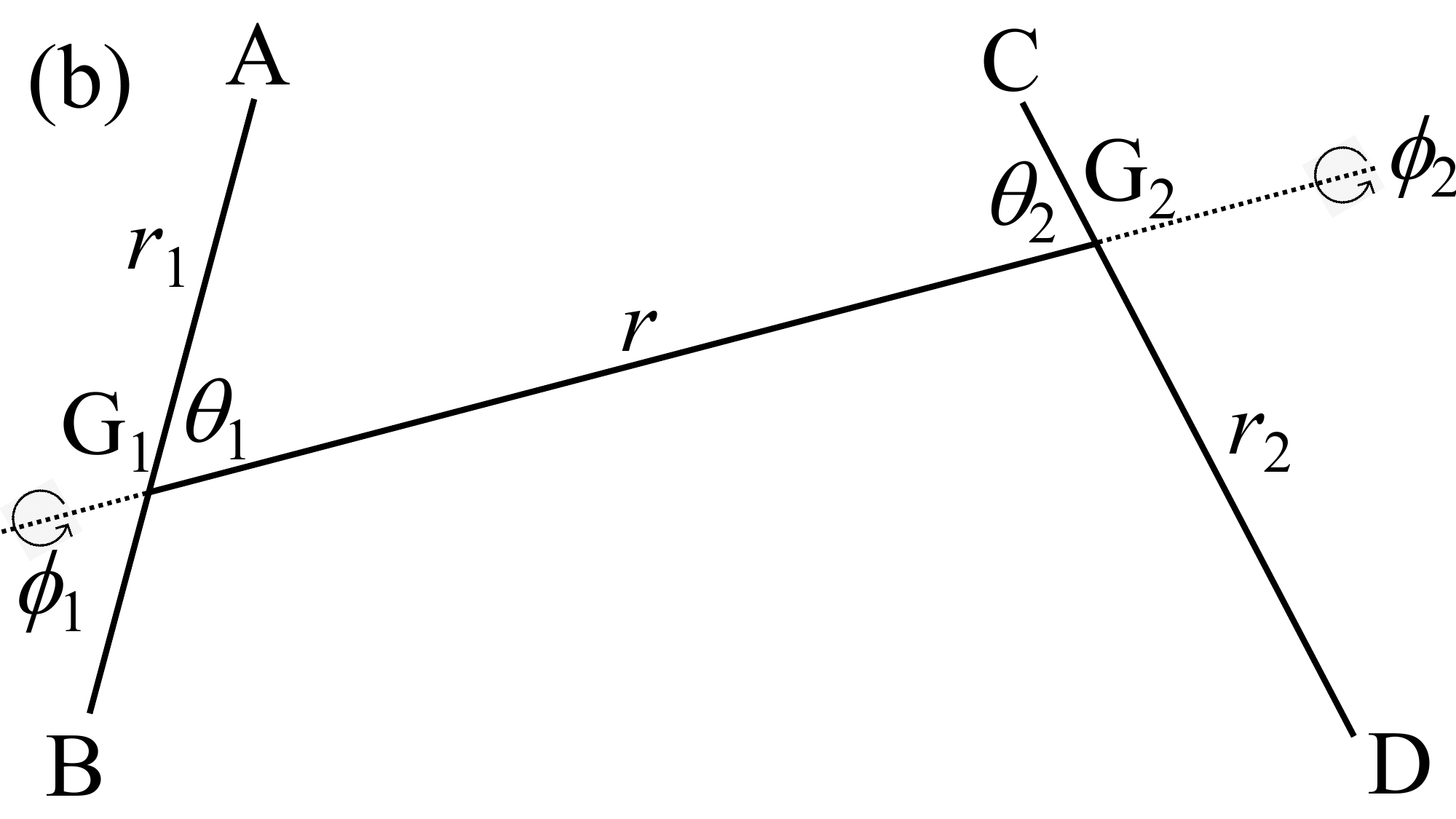} &&
\includegraphics[width=8cm]{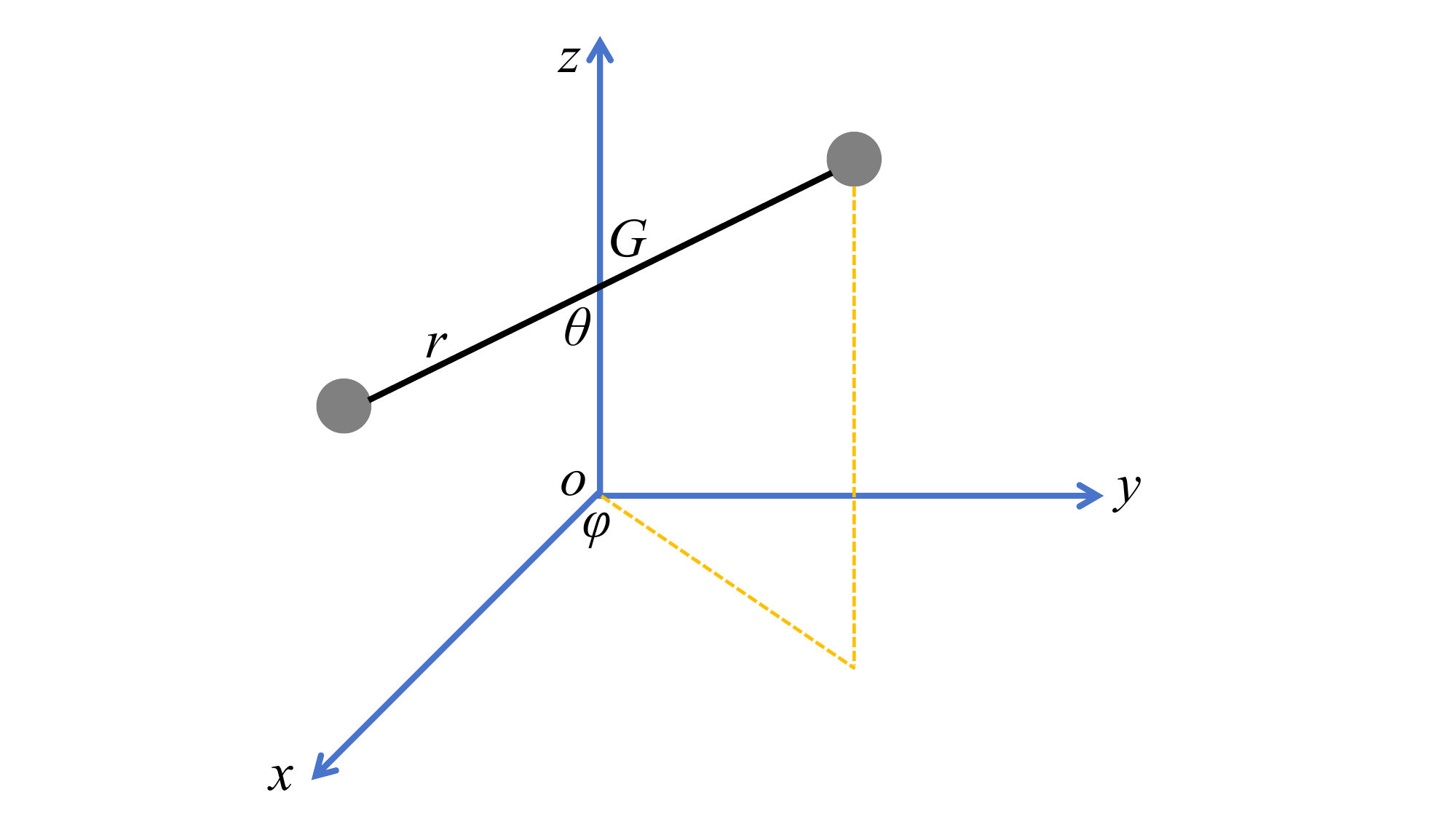}  \\
2 && Coordinates&& $\{r_{\mathrm{d}},r_{\mathrm{v}},\theta\}$ &&
$\{r,r_1,r_2,\theta_1,\theta_2,\phi=\phi_1-\phi_2\}$ &&
$\{x,y,z,r,\theta,\phi\}$ \\
3 && Tolerance error && $10^{-4}$ && $10^{-4}$ && $10^{-4}$ \\
4 && Validate error  && $\sim10^{-3}$ && $\sim10^{-2}$ && $\sim10^{-2}$  \\
5 && $n$ && $9.0\times10^3$ && $9.5\times10^4$ && $9.5\times10^4$ \\
  \hline
 \end{tabular}
  \end{center}
   \label{tab:examples}   
    \end{sidewaystable}

\clearpage
\begin{table}
 \caption{%
DVR-grids of the wave function representation and parameters of the
initial condition for the present MCTDH calculations. The definitions
of the coordinates (indicated in the first column) are given in Table
\ref{tab:examples}. The
second column describes the primitive basis functions, which underlay
the DVR basis. The third column gives the number of the grid points.
The fourth column gives
the range of the grids in bohr or radian. The fifth column gives the
symbol of the one-dimensional (1D) function for each coordinate of the
initial wave function in relaxations for the ro-vibrational eigen-states.
The other columns give the parameters for these 1D functions, including
positions and momenta in the 1D function, width of the Gaussian function
({\it i.e.}, variance of the modulus-square of the Gaussian function,
$W_{\mathrm{GAUSS}}$), and the initial quantum number $j_{\mathrm{ini}}$
of the angular function.
}
\begin{center}
 \begin{tabular}{llllllrlrlrlrlr}
  \hline
Coordinates &~~& \multicolumn{5}{c}{Primitive basis function}
&~~~~~~& \multicolumn{7}{c}{Initial wave function}  \\ \cline{3-7} \cline{9-15}
&~~& Symbol $^a$ &~~& Grid points &~~& Range of the grids 
&~~~~~~& Symbol $^b$ &~~& Position &~~& Momentum &~~& Parameters  \\
\hline
$r_d$    && SIN && $68$ && $[1.00,9.04]$    && GAUSS && $4.50$ && $-8.00$ &&
$W_{\mathrm{GAUSS}}=0.25$ Bohr \\
$r_v$    && SIN  && $48$  && $[0.60,6.24]$  && EIGENF&& --    && --      && ground state \\
$\theta$ && LEG  && $31$  && $[0,\pi]$      && LEG   && --    && --      && 
$j_{\mathrm{ini}}=0$, sym \\
\hline
 \end{tabular}
  \end{center}
   \label{tab:numerical-details}
    \end{table}
\quad  \\
$^a$ Symbol ``SIN'' stands for sine DVR. Symbol ``LEG'' denotes one-dimensional
Legendre DVR for angular coordinate. \\
$^b$ Symbol ``GAUSS'' designates the choice of Gaussian function as
initial SPFs. Symbol ``EIGENF'' means eigenfunction of a specified
potential which, in this work, is reduced one-dimensional potential
function of H$_2$ calculated from the PES. Symbol ``LEG'' denotes
Legendre function to specify the initial wave function.

\clearpage
 \section*{Figure Captions}
  
\figcaption{fig:methods}{%
Construction schemes of the potential function in the CPD form
where the training data have already been computed by {\it ab initio}.
This special expansion tensor form is useful and helpful to resolving
high-dimension working equations of quantum dynamics. The black boxes
give computational inputs and outputs. The red boxes give computational
techniques. Here, the regression methods contain those of either NN or
GPR, while the re-fitting methods mean POTFIT, MGPF, MLPF, MCPF, or
MCCPD. The clockwise pathway represents the traditional scheme reproducing
the PES in an arbitrary form. In general, this form can be used for both
classical and quantum dynamics calculations. For special cases, such as
the ML-MCTDH calculations, the PES has to be re-fitted in either SOP or
CPD form. The anticlockwise pathway represents the direct scheme proposed
by the previous CPD-GPR and the present CPD-SVR approaches. This scheme
directly reproduces the PES in the CPD form through either GPR or SVR.
Therefore, the traditional and direct schemes have two and one regression
steps, respectively. Since numerical regressions reproduce errors in
either traditional or direct scheme, the direct scheme with only one
step is much more efficient and accurate than the traditional one.
}%

\figcaption{fig:svr-prin-00}{%
Schematic illustrations on (a) the core ideas of SVR and support vector
and (b) the dependance of the tolerance error and the fitting error,
where the abscissa and ordinate axes represent the coordinate $\mathbf{x}$
and electronic energy $V$, respectively. The gray
line represents the target PES denoted by $0$, while the red and blue
lines are parallel to the prediction and denoted by $\pm\epsilon$. In
SVR, only the points outside the region surrounded by red and blue lines
contribute to the cost insofar. It turns out that in most cases the
optimization problem can be solved more easily in its dual formulation.
The support vectors in SVR are the data that lie on and outside the two
hypersurfaces (represented by blue and red lines) that are parallel to
the target hypersurface, namely the SVR prediction (represented by gray
line) with distances of tolerance error $\pm\epsilon$. The support vectors
are points most difficult to regress such that they have direct bearing
on the optimum location of the target prediction function. The optimal
PES (the gray line) stems from the function class with the biggest
``capacity'' of independent samples that one can twiddle keeping the
functional form not too complex. Furthermore, in subfigure (b) the
solid and dashed lines represent the hypersurfaces for $\epsilon$ and
$\epsilon'$, respectively, while the blue and red lines are those
associated with those with different symbols. The
optimal potential function (the gray line) stems from the function
class with the biggest capacity of independent samples that one can
twiddle keeping the functional form not too complex. Therefore, in
principle, the fitting error depends on the tolerance error. If only
few of noises exist in the training database the support vectors
should independ on the choice of tolerance error.
}%
 
\figcaption{fig:alg}{%
Updating scheme of the classifications of the training points when a
new point is added into the optimization procedure. The core idea of
SVR and support vector is given in the main text. When a new training point is added, the decision
hypersurface should be updated so that all of previously optimized
results should be updated according to the present ws-SVR approach.
The upper and lower panels illustrate two cases for adding new training points.
Given in the upper panel is Case \#1, where the new training point
is added into $\mathcal{R}$ implying that $\vert h\vert<\epsilon$
and the other points are unchanged. The lower panel illustrates
Case \#2, where the new training point is added into $\mathcal{E}$
or $\mathcal{S}$. In Case \#2 the decision surface of the last round
has to be changed and hence the classifications are changed. To keep
the training points belonging to their original sets as much as possible,
one can find four different situations in changing the classification
during optimization. In these subfigures, the red, blue, and yellow
black circles represent the training points belonging to $\mathcal{S}$,
$\mathcal{R}$, and $\mathcal{E}$, respectively, while the black point
represents the newly adding point.
}%

\figcaption{fig:conver}{%
Validation errors (RMSE values in hartree) of the present CPD-ws-SVR
calculations as functions of number of training data $n$ for the
H + H$_2$ system. These SVR calculations are performed using various
tolerance errors. The red, blue, and black lines and symbols represent
errors computed with tolerance errors of $10^{-2}$, $10^{-3}$,
and $10^{-4}$ hartree, respectively. All of training data are
computed through pre-existed BKMP2 potential function where they are
randomly sampled in the whole configurational space. The error bars
are obtained through $5$-fold cross validation ($5$-CV), where the
validation set comprises of $20\%$ of the whole database.
}%

\figcaption{fig:supp-vec}{%
Numbers of training data $n$ and support vectors $n_{\mathrm{sv}}$
together with ratios $n_{\mathrm{sv}}/n$ for various potential energy
intervals of the H + H$_2$. In fitting this PES, the tolerance error
is $10^{-4}$ hartree. The abscissa axis gives seven uniform intervals
for the potential energy less than zero. Each interval represents an
energy range of about $0.62$ eV. The left axis represents numbers of
data while the right axis represents the ratio $n_{\mathrm{sv}}/n$.
The red and blue columns give numbers of support vectors and training
data, respectively. The black dots represent the ratios $n_{\mathrm{sv}}/n$
in percentage. The support vectors in SVR are the data that lie on and
outside the two hypersurfaces that are parallel to the target hypersurface.
Thus, the
support vectors are points most difficult to regress such that they
have direct bearing on the optimum location of the target prediction
function. In this context, ratios $n_{\mathrm{sv}}/n$ must be smaller
than one.
}%

\figcaption{fig:pes-contor}{
The contour cuts of the SVR potential tensor on the $r_{\mathrm d}$-$r_{\mathrm v}$
plane with $\theta=\pi/2$ for the H + H$_2$ system. The definitions of
coordinates $r_{\mathrm v}$, $r_{\mathrm d}$, and $\theta$ are given
in Table \ref{tab:examples}. Here, we set the potential
energy of the totally separated three H atoms to be zero. To draw this
potential tensor, the grids of the $r_{\mathrm d}$-$r_{\mathrm v}$ plane
are set to be $68\times48$ equal to those given in Table \ref{tab:numerical-details}.
Moreover, the 3D graphics of the potential function $V(r_{\mathrm d},r_{\mathrm v},\theta=\pi/2)$
is also given, where the coordinate variables $r_{\mathrm d}$ and $r_{\mathrm v}$
are given in bohr while the potential energy is given in eV.
}

\begin{figure}[h!]
  \centering
   \includegraphics[width=16cm]{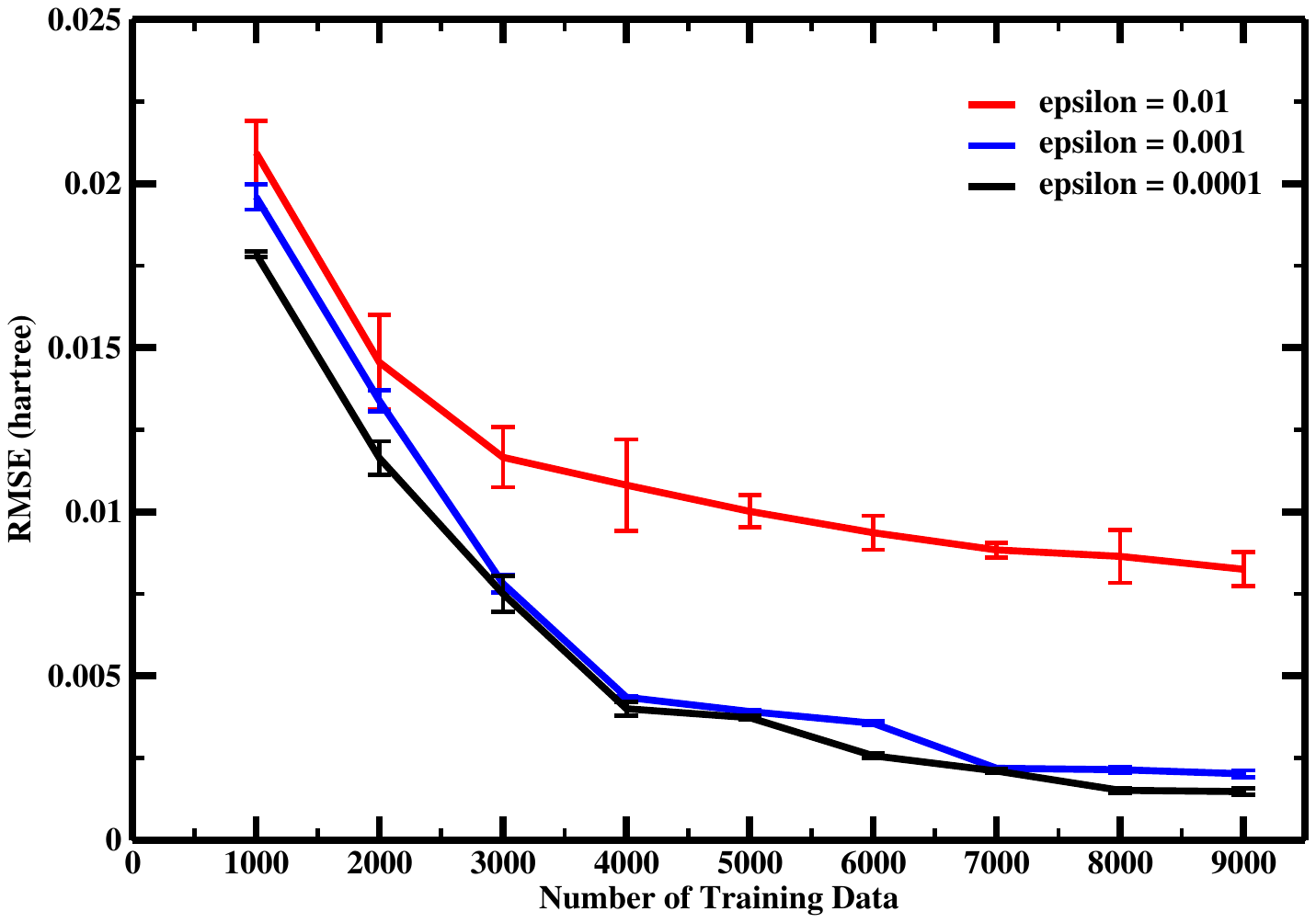}
    \caption{\figfoot}
     \label{fig:conver}
      \end{figure}

\figcaption{fig:flux}{%
Energy-dependent reactive probabilities versus total energy (in eV)
for the H' + H$_2$ $\to$ HH' + H reaction where the PES is trandsferred
into various tensor forms in performing the quantum dynamics calculations.
The green, blue, red,
and black lines represent the reactive probabilities computed on the
basis of the {\it potfited}, {\it mccpded}, CPD-GPR, and CPD-ws-SVR
potential tensors. The {\it potfitted} and {\it mccpded} potential
tensors are constructed by POTFIT and MCCPD through the BKMP2 PES,
respectively. Numerical details of these calculations were reported
in Reference \cite{son22:11128}. The CPD-GPR and CPD-ws-SVR tensors
are constructed on the basis of the same training sets with $n=9000$,
where the CPD-GPR calculation was already reported in Reference
\cite{son22:11128}. Error convergence of the CPD-ws-SVR calculation is illustrated
in Figure \ref{fig:conver}.
}%

\figcaption{fig:svr-con}{%
Validation errors (RMSE values in hartree) of (a) the present CPD-GPR calculations
and (b) the present 6D CDP-ws-SVR calculations as functions of number of training
data $n$ for H + H$_2$, H$_2$ + H$_2$, and H$_2$/Cu(111). The blue and red lines
represent errors computed for the H$_2$ + H$_2$ and H$_2$/Cu(111) systems,
respectively, while the green line in subfigure (a) gives errors for H + H$_2$.
The similar CPD-ws-SVR error curves for H + H$_2$ are illustrated in Figure \ref{fig:conver}.
The errors are obtained by averaging five-times calculations at the same
setup, where the validation set comprises of $20\%$ of the whole database.
All of data, in either training set or validation set, are randomly sampled
in the configurational space.
}%

\begin{figure}[h!]
  \centering
   \includegraphics[width=16cm]{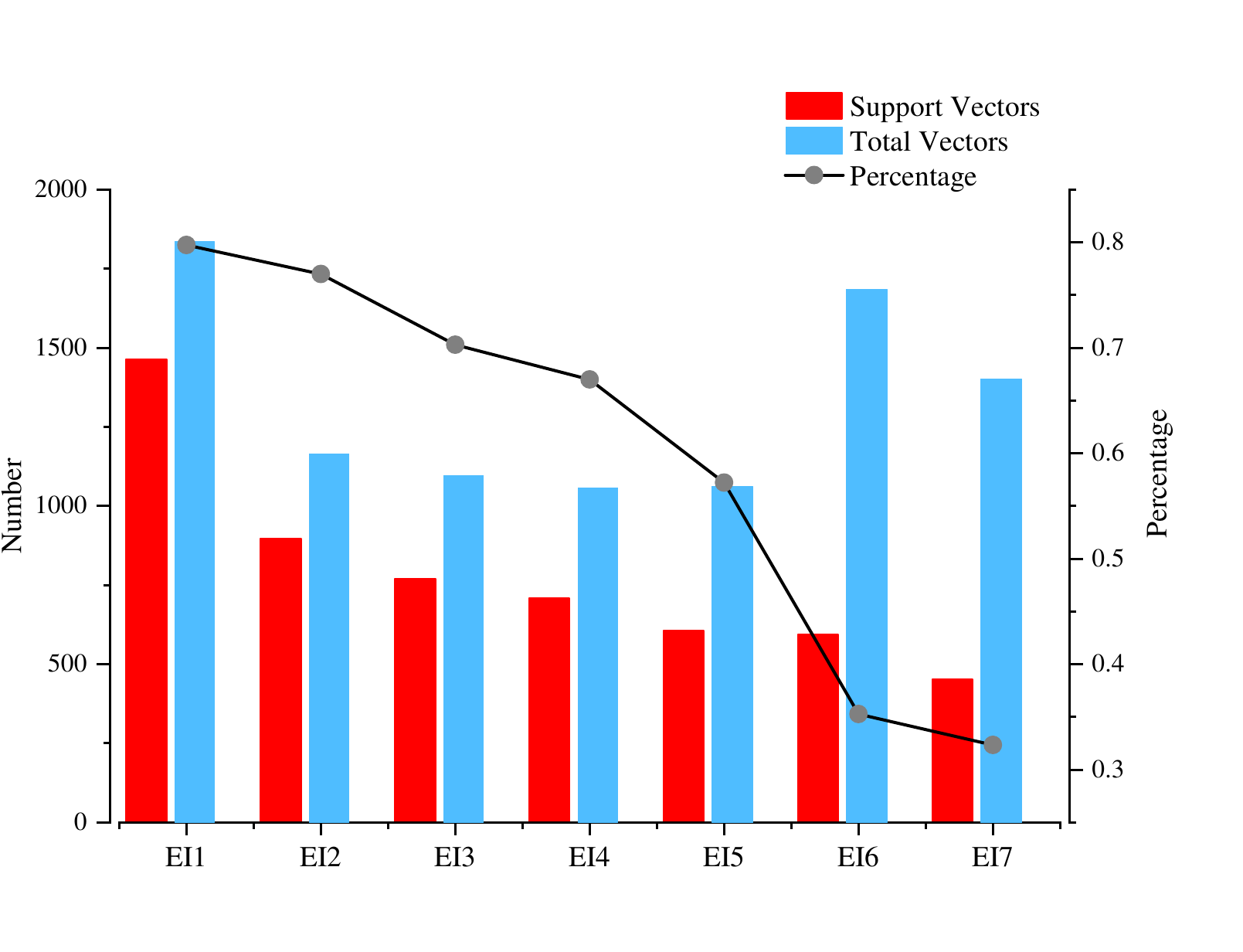}
    \caption{\figfoot}
     \label{fig:supp-vec}
      \end{figure}

\figcaption{fig:ration-sv}{%
The ratios $n_{\mathrm{sv}}/n$ as function of the number of training
data $n$ for the H$_2$ + H$_2$ and H$_2$/Cu(111) systems, given by
the blue and red lines, respectively, where $n_{\mathrm{sv}}$ is the
number of support vectors. The same curve for the H + H$_2$ system has
been illustrated by the black dots of Figure \ref{fig:supp-vec}. The
abscissa and ordinate axes represent number of training data $n$ and
ratios $n_{\mathrm{sv}}/n$, respectively. For simple, the abscissa
axis is given by logarithmic scale instead of linear scale.
}%

\clearpage

\clearpage
 \begin{figure}[h!]
  \centering
   \subfigure[\quad Core ideas of SVR and support vector]{%
    \includegraphics[width=16cm]{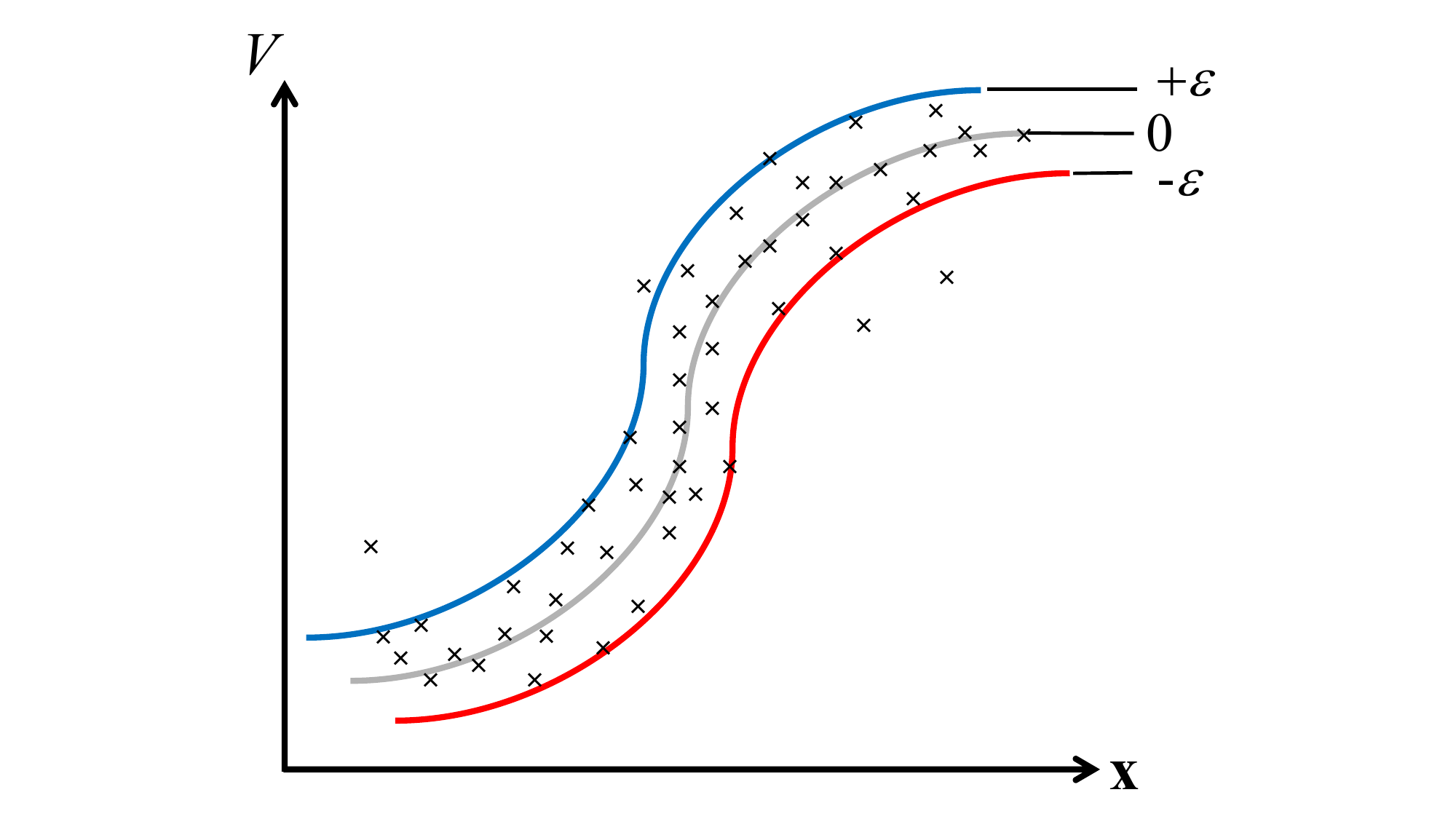}
     }%
      \end{figure}

\clearpage
      
\clearpage

\clearpage

\clearpage

\clearpage

\clearpage
 \begin{figure}[h!]
  \centering
   \subfigure[\quad Convergence inspection of the CPD-GPR calculations]{%
    \includegraphics[width=16cm]{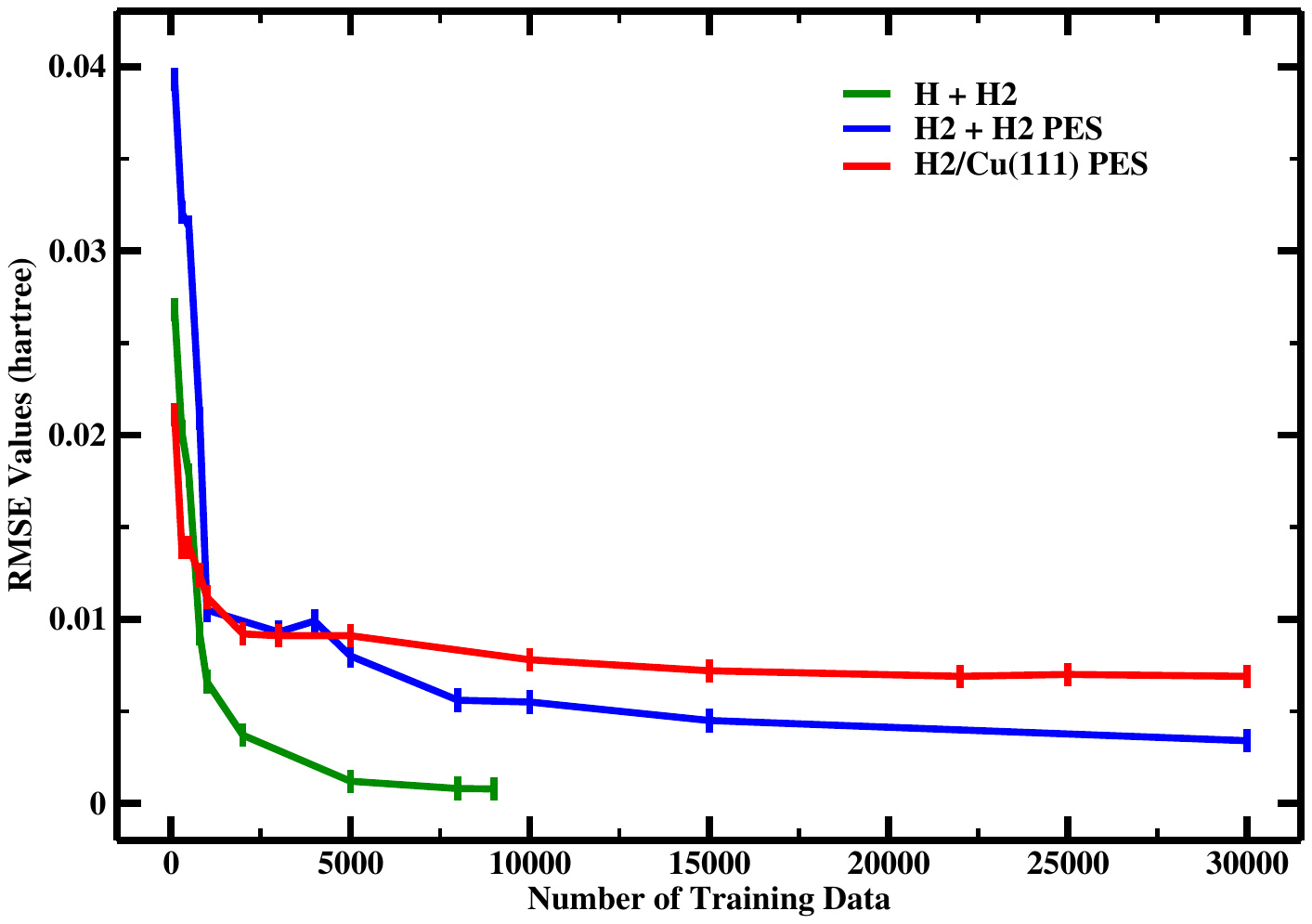}
     }%
     \end{figure}
\begin{figure}[h!]
 \centering
  \subfigure[\quad Convergence inspection of the CPD-ws-SVR calculations]{%
   \includegraphics[width=16cm]{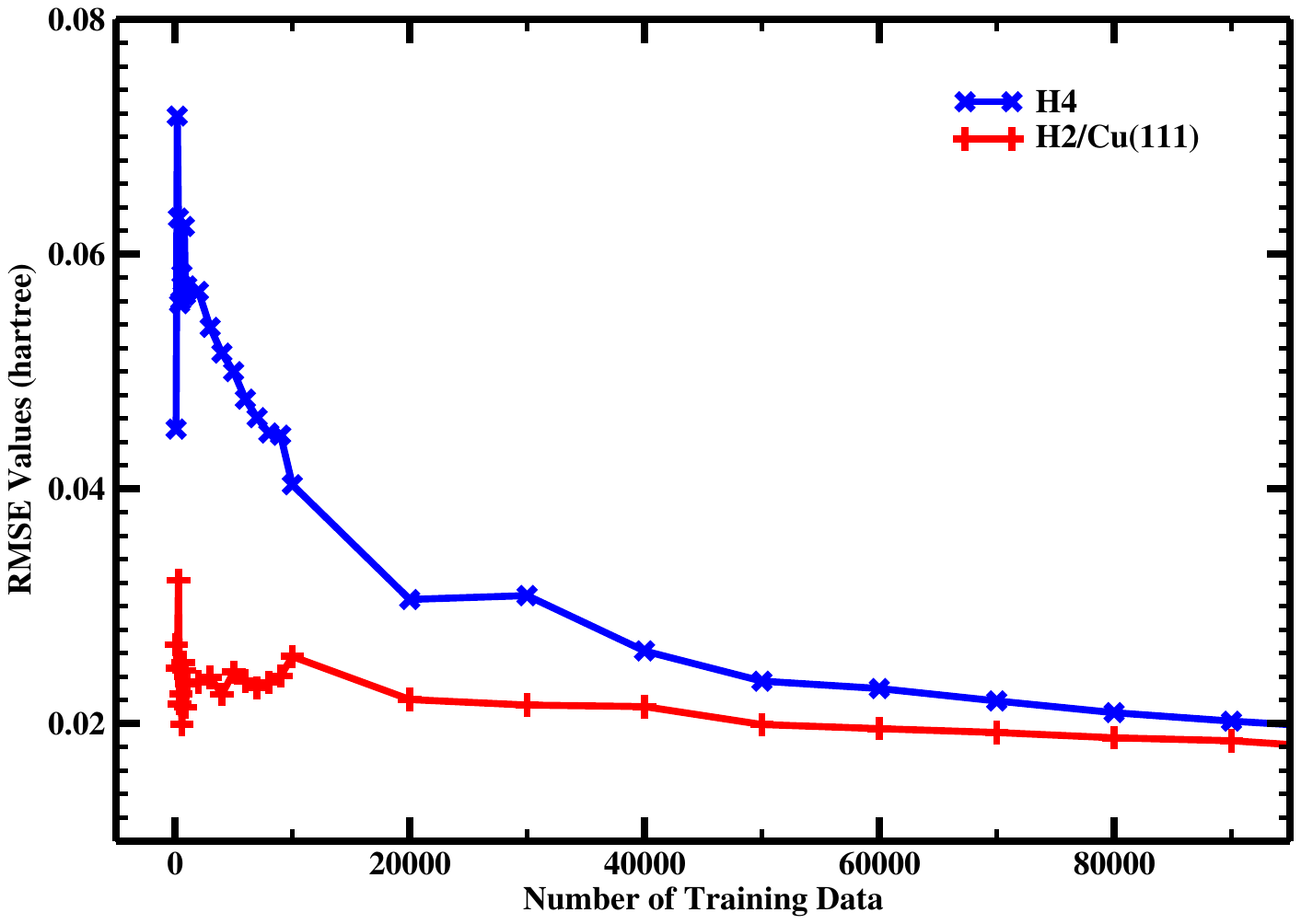}
    }%
    \caption{\figfoot}
     \label{fig:svr-con}
      \end{figure}
                       
\clearpage
      
\end{document}